%% file: main.tex
\documentclass[english,journal=jctcce,manuscript=article,email=true,etalmode=truncate,maxauthors=0]{achemso}

\input{preamble}

  \author{Karl Pierce}
  \affiliation{ Center for Computational Quantum Physics, Flatiron Institute, 162 5th Ave., New York, 10010  NY,  USA}
  \email{kpierce@flatironinstute.org}

   \author{Miguel Morales}
  \affiliation{ Center for Computational Quantum Physics, Flatiron Institute, 162 5th Ave., New York, 10010  NY,  USA}

  \title{Using Matrix-Free Tensor-Network Optimizations to Construct a Reduced-Scaling and Robust Second-Order M{\o}ller-Plesset Theory}

\begin{document}

\begin{abstract}
We investigate the efficient combination of the canonical polyadic decomposition (CPD) and tensor hyper-contraction (THC) approaches. 
We first present a novel low-cost CPD solver which leverages a precomputed THC factorization of an order-$4$ tensor to efficiently optimize the order-$4$ CPD with $\mathcal{O}(NR^2)$ scaling.
With the matrix-free THC-based optimization strategy in hand we can: efficiently generate CPD factorizations of the order-4 two-electron integral tensors; and develop novel electronic structure methods which take advantage of both the THC and CPD approximations.
Next, we investigate the application of a combined CPD and THC approximation of the Laplace transform (LT) second-order M{\o}ller-Plesset (MP2) method.
We exploit the ability to switch efficiently between the THC and CPD factorizations of the two electron integrals to reduce the computational complexity of the LT MP2 method while preserving the accuracy of the approach.
Furthermore we take advantage of the robust fitting approximation to eliminate leading order error in the CPD approximated tensor networks.
Finally, we show that modest values of THC and CPD rank preserve the accuracy of the LT MP2  method and that this CPD+THC LT MP2 strategy realizes a performance advantage over canonical LT MP2 in both computational wall-times and memory resource requirements.
\end{abstract}



\section{Introduction}
\begin{figure}[t]
        \includegraphics[trim=5cm 5cm 5cm 5cm, clip, width=0.7\columnwidth]{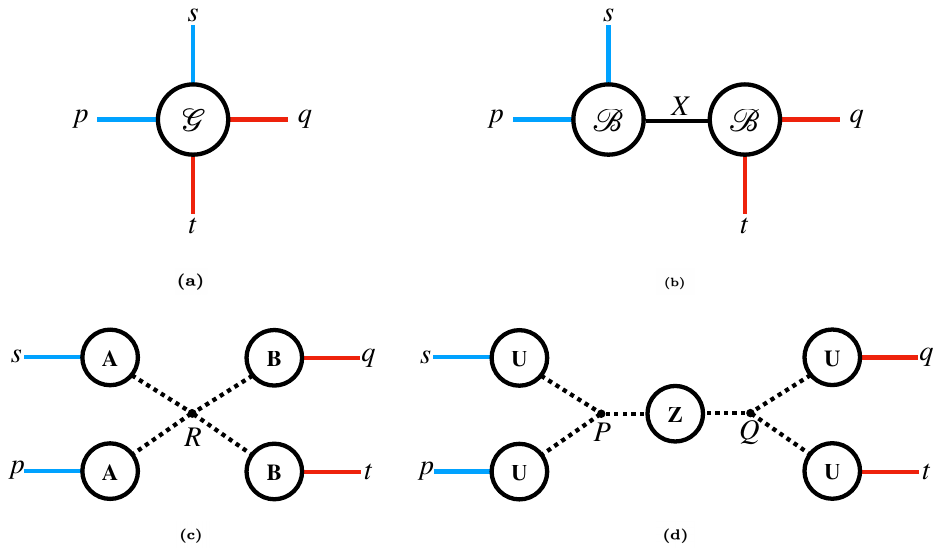}
    \caption{Representations of the order-4 two-electron integral tensor (a). (b) shows the density fitting approximation of $\mathcal{G}$,(c) shows the canonical polyadic decomposition of $\mathcal{G}$, and (d) shows the tensor hyper-contraction approximation of $\mathcal{G}$.}
    \label{fig:tensor-diagram}
\end{figure}

In the search for low-scaling and accurate electronic structure methods, chemists have found success in the application of low-rank tensor decompositions to the hierarchy of wavefunction-based methods.
Among the most important low-rank approximations in electronic structure methods has been the density fitting (DF) approximation\cite{VRG:whitten:1973:JCP,VRG:dunlap:1979:JCP,Vahtras:1993:CPL,VRG:jung:2005:PNAS} of the two-electron integral tensor (TEI), $\mathcal{G}$.
The TEI tensor and the DF approximation of the TEI tensor are represented in \cref{fig:tensor-diagram}a and \cref{fig:tensor-diagram}b, respectively.
The DF approach has been quickly and successfully adopted by the community because of its low-cost, robust construction, via the introduction of predefined auxiliary basis sets, and its capacity to reduce the storage complexity of the TEI tensors.
However, it is well known that the DF approximation is unable to reduce the computational complexity of high-cost bottleneck tensor contractions in all levels of electronic structure methods.

In an effort to reduce the computational scaling of these high-cost tensor contractions, chemists have studied the application of other low-rank tensor formats. 
In this work, we focus on the application of the 
tensor hyper-contraction\cite{VRG:hohenstein:2012:JCP,VRG:hohenstein:2012:JCPa,VRG:parrish:2012:JCP,Hohenstein:2013:JCP,VRG:parrish:2014:JCP,Shenvi:2013:JCP,Schutski:2017:JCP,Parrish:2019:JCP,Lee:2019:JCTC,VRG:hummel:2017:JCP,VRG:song::JCP,Hohenstein:2019:JCP,Hohenstein:2021:JCP,Hohenstein:2022:JCP,Jiang:2022:JCTC,Zhao:2023:JCTC,Datar:2024:JCTC} (THC) approximation, see \cref{fig:tensor-diagram}d.
The THC approximation has gained popularity because it efficiently preserves the accuracy of electronic structure methods and it can be constructed efficiently.
Specifically, there have been many recent studies on the application of interpolative separable density fitting\cite{Lee:2019:JCTC, VRG:hu:2017:JCTC,VRG:lu:2015:JCP,Matthews:2020:JCTC} (ISDF) approximation to initialize the THC decomposition.\footnote{The ISDF has been particularly useful for periodic systems for which it can be used to constructed the THC in $\mathcal{O}(N^3)$ time. For molecules there is currently no implementation which reduces the cost to $\mathcal{O}(N^3)$, though we are aware of multiple efforts which look to make this improvement.
In this work, we do not focus on the construction of the THC decomposition and instead assume there exists some algorithm to efficiently construct it.}
The ISDF based THC approximation has been used gainfully to reduce the cost of algorithms in both molecular (MP2 \cite{Lee:2019:JCTC}, GW \cite{ISDF_GW_Duchemin21}) and periodic (RPA \cite{ISDF_RPA_Yeh2024}, GF2 \cite{ISDF_GF2_Pokhilko_2024}, GW \cite{ISDF_GW_Yeh2024} and Auxiliary Field Quantum Monte Carlo \cite{ISDF_AFQMC_Malone19}) calculations.
Unfortunately, much like the DF approximation, the THC representation is limited in its ability to formally reduce the complexity of tensor networks which are contracted across indices of different particles.

The canonical polyadic decomposition (CPD) of the order-4 TEI tensor (\cref{fig:tensor-diagram}c) has been shown to be more flexible in its ability to reduce these high-cost tensor-networks\cite {VRG:chinnamsetty:2007:JCP,Benedikt:2013:MP,Benedikt:2013:JCP,Bohm:2016:JCP,Schutski:2017:JCP}.
Though the CPD format is incredibly powerful, the costs associated with its construction are prohibitively large.
A typical analytic CPD optimization carries an iterative complexity of $\mathcal{O}(N^4R_\mathrm{CPD})$ where $N$ is the dimension of the target order-$4$ tensor and $R_\mathrm{CPD}$ is the chosen CP rank.
In our previous study,\cite{VRG:pierce:2023:JCTC} we show that it is possible to leverage the density fitting approximation to reduce the complexity of the alternating least squares minimization of the order-4 Coulomb integral tensor from an $\mathcal{O}(N^4R)$ scaling and an $\mathcal{O}(N^4)$ storage cost to an $\mathcal{O}(N^3R_\mathrm{CPD})$ scaling and an $\mathcal{O}(N^3)$ storage cost, where $R_\mathrm{CPD}$ is the desired CPD rank.\footnote{Please note that here $R_\mathrm{CPD}$ does not mean the exact CP rank of the problem, as there is no finite algorithm to compute the exact CP rank of a given tensor.\cite{Hastad:1990:algorithm,VRG:hillar:2013:JA}}
Even with this formulation, the optimization of the order-4 CPD can still be expensive.
Therefore, the efficient construction and application of the CPD approximated order-4 TEI to electronic structure methods is an open direction of research.
In an effort to extend the effective application of the CPD of high-order tensors, we introduce a novel alternating least-squares algorithm that efficiently leverages the THC approximation.
We take advantage of the THC format to compute approximate, matrix-free gradients of the CPD least-squares loss function.
With this algorithm, we reduce the operational and storage complexity of the CPD optimization.


Subsequently, we develop a novel MP2 strategy to effectively combine the high-accuracy THC decomposition with the algorithmically flexible CPD representation.
We integrate the CPD into an THC-based MP2 algorithm first developed by Hohenstein et al.\cite{VRG:hohenstein:2012:JCP,VRG:kokkilaschumacher:2015:JCTC} and recently studied using the ISDF by Lee et al.\cite{Lee:2019:JCTC}
With the both the CPD and THC approximations in hand, we have the freedom to choose when and how to apply these approximations.
In the MP2 algorithm, we decide to only to introduce the CPD into the the exchange component of the MP2 energy.
Furthermore, we utilize the robust tensor approximation\cite{VRG:dunlap:2000:PCCPP, VRG:pierce:2021:JCTC} to eliminate the leading-order error associated with the CPD.
We demonstrate how this careful strategy requires modest values of THC and CPD ranks to compute relevant energies to a high degree of accuracy.
Most dramatically we highlight that this new strategy reduces the memory overhead of the MP2 approach from $\mathcal{O}(N^4)$ to $\mathcal{O}(R^2) \approx \mathcal{O}(N^2)$ where $R$ is the THC or CP rank.

The remainder of the manuscript is organized as follows. \Cref{sec:formal} introduces the low-rank tensor decompositions described above, demonstrates how the THC can be leveraged in the construction of the CPD and walks through the robust and low scaling tensor approximated MP2 algorithm. In \cref{sec:comp} we describe our computational experiments and in \cref{sec:results} we compare our low-scaling MP2 algorithm to the standard and the THC-approximated Laplace transform (LT) MP2 algorithm. Finally, Our findings are summarized and we discuss future directions of research in \cref{sec:conclusions}.

\section{Formalism}\label{sec:formal}
\subsection{Low-Rank Tensor Approximations}
Here we will focus on the decomposition of the two-electron integral (TEI) tensor as a means to reduce the complexity of the canonical second-order M{\o}ller-Plesset perturbation theory algorithm.
The TEI tensor can be expressed in a generic set of single-particle basis functions as 
\begin{align}\label{eq:g}
    g^{pq}_{st} = \iint \phi^*_{p}(r_1) \phi^*_q(r_2) g(r_1, r_2) \phi_s(r_1) \phi_t(r_2) dr_1 dr_2
\end{align}
where the kernel $g(r_1, r_2)$ is positive definite; in this manuscript we only consider the Coulomb interaction kernel $g(r_1, r_2) = |r_1 - r_2|^{-1}$.
Please notice that stacked index labels, such as $p$ and $s$, denote functions of a single particle.

\begin{figure}[t]
        \includegraphics[trim=1cm 8cm 1cm 9cm, clip, width=\columnwidth]{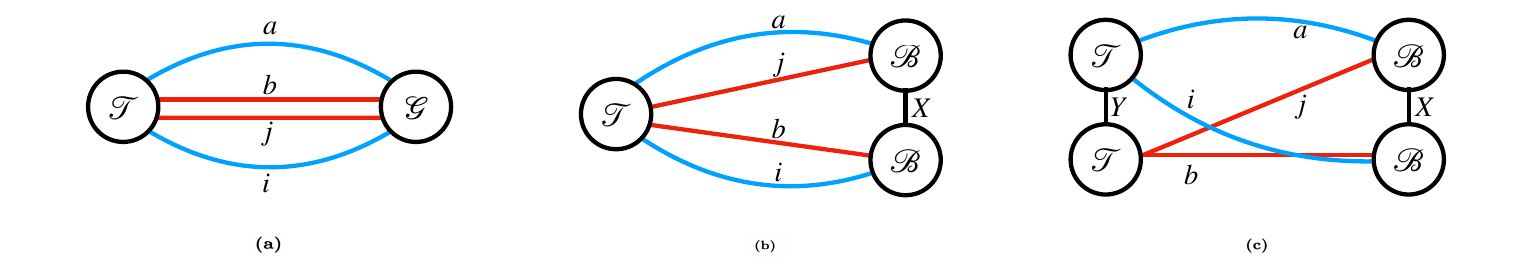}
    \caption{The LT MP2 correlation energy tensor-network. (a) utilizes the canonical order-4 TEI for both tensors in the network, (b) uses a mixture of the canonical order-4 tensor and the DF approximation of the TEI tensor, and (c) uses the DF approximation of the TEI for both tensors in the network.}
    \label{fig:df-contract}
\end{figure}

The most common approximation of the TEI tensor is the density fitting (DF) approximation. 
The DF approximation represent the TEI tensor in the following form
\begin{align}\label{eq:SQG}
    g^{pq}_{st} \overset{\mathrm{DF}}{\approx} \sum_X B^{pX}_{s} B^{qX}_{t}
\end{align}
where $B$ is referred to as an order-3 DF integral tensor.
We can conveniently form the DF approximation via an optimized, predetermined set of auxiliary functions, denoted as $X$. 
One can see now that the DF approximation pins both functions of a single particle to the same DF integral tensor.
With this topological constraint, the DF approximation is unable to reduce the complexity of tensor contractions involving indices associated with different particles.
This restriction is demonstrated in \cref{fig:df-contract}, i.e. all the tensor-networks in this figure formally have a computational scaling of no less than $\mathcal{O}(N^4)$.

The THC approximation of the order-4 TEI tensor is defined as
\begin{align}\label{eq:THC}
    \bar{g}^{pq}_{st} = \sum_{PQ} U^p_P U^P_s Z^P_Q U^q_Q U^Q_t \overset{\mathrm{THC}}{\approx} g^{pq}_{st}
\end{align}
where the dimensions $P$ and $Q$ refer to the THC rank, $R_\mathrm{THC}$, and in this study $R_\mathrm{THC} = P = Q$.
The THC rank is called a `hyper-index` because more than 2 tensors carry the index.
We denote the the THC approximation of a tensor with a bar over the tensor.
The tensors $\bf U$ are the THC factor matrices and the tensor $\bf Z$ is the THC core tensor.
In \cref{eq:THC}, all of the THC factor matrices use the same label $\bf U$ because we assume $\{p,q,s,t\}$ are labels of the same orbital basis set and, therefore, each $\bf U$ is identical.
If we assume the label $p$ represents the complete molecular orbital basis, we can then partition $U$ into two components ${\bf U} = \{{\bf U}^o, {\bf U}^v\}$ where $o$ represents the occupied orbital basis set and $v$ represents the unoccupied orbital basis set.

\begin{figure}[b]
        \includegraphics[trim=5cm 7cm 5cm 7cm, clip, width=.7\columnwidth]{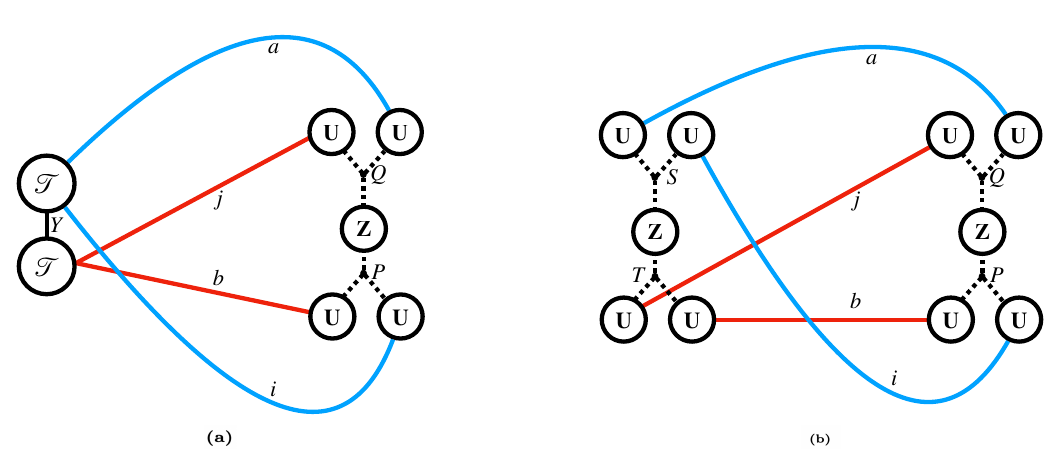}
    \caption{The LT MP2 correlation energy tensor-network. (a) utilizes a mixture of the DF approximation of the TEI tensor and the THC approximation of the TEI tensor and (b) utilizes the THC approximation for both tensors in the network.}
    \label{fig:thc-contract}
\end{figure}

The THC approximation efficiently decomposes the order-4 TEI tensor into a set of order-2 tensors. 
This approximation decouples all of the original indices of the TEI tensor and reduces the storage complexity of the tensor from $\mathcal{O}(N^4)$ to $\mathcal{O}(N^2)$.
This $\mathcal{O}(N^2)$ storage scaling assumes the THC rank grows linearly with system size, though the exact scaling of the THC rank is still an open question and is understood to vary with tensor basis, for example $R_\mathrm{THC}(\bar{g}^{ab}_{ij}) < R_\mathrm{THC}(\bar{g}^{pq}_{st})$.
The THC can be initialized relatively efficiently; iterative and non-iterative methods have been subject of other studies into the THC.
As of today, molecular THC optimization strategies have a formal computational complexity no less than $\mathcal{O}(N^4)$.
Though, we and other researcher groups are currently investigating methods to reduce the scaling of the THC optimization.
As one can see from \cref{eq:THC}, the THC approximation pins indices of a single particle to the same hyper-index.
Therefore, this representation has topological limitations which are similar to the DF approximation, best illustrated by \cref{fig:thc-contract}.
The tensor-networks in \cref{fig:thc-contract} are approximations of \cref{fig:df-contract}a.
Unfortunately, the contraction of either of the the THC networks in \cref{fig:thc-contract} still has a computational complexity of $\mathcal{O}(N^4)$. 
As a means to reduce the complexity of terms like \cref{fig:df-contract}a, researchers have studied the application of the order-4 canonical polyadic decomposition of the TEI tensor.

The CPD\cite{VRG:carroll:1970:P,Harshman:1970:WPP} of the order-4 TEI tensor can be written as the following\footnote{A full review of the CPD will not be provided in this work; if interested we direct readers to the review by Kolda et. al.\cite{VRG:kolda:2009:SR}} 
\begin{align}\label{eq:cp4}
    \tilde{g}^{pq}_{st}(\vec{x}) = \sum_r^{R_\mathrm{CPD}} \lambda_r A^p_r A^r_s B^q_r B^r_t
\end{align}
where ${\bf A,\bf B}$ are CPD factor matrices, $\vec{x} = [\bf{A}, \bf{B}, \bf \lambda]$, $R_\mathrm{CPD}$ is the CP rank, and $\lambda_r$ is a scaling coefficient which enforces the unit normalization of the CPD factor matrices. 
Please notice that the CPD approximation of a tensor will be denoted with a tilde over the tensor and, without loss of generality, $\bf \lambda$ can be factored into one or many of CPD factor matrices and will therefore be ignored in the further equations.
In \cref{eq:cp4}, one can see that the CPD factors of particle 1 are represented with the matrix $\bf A$ and factors of particle 2 are represented with the matrix $\bf B$. 
We denote them in this way because we do not force the tensors of particle 1 to be equivalent to particle 2.
In principle, this freedom can lead to broken symmetry approximations of $\mathcal{G}$.
However, in general, we find the degree of symmetry breaking is small and decreases with increasing $R_\mathrm{CPD}$ which is consistent with findings from Hackbush et al.\cite{VRG:benedikt:2011:JCP}
We, again, assume the indices $\{p,q,s,t\}$ span the same basis and therefore both $\bf A$ matrices and $\bf B$ are essentially equivalent. If we assume, again, that $p$ is the complete molecular orbital basis, we can also partition these CPD factors into occupied and unoccupied components, i.e. $\bf A = \{A^o, A^v\}$.
And, like the THC approximation, we find that the CPD rank of a tensor depends on the tensors orbital indices, for example $R_\mathrm{CPD}(\tilde{g}^{ab}_{ij}) < R_\mathrm{CPD}(\tilde{g}^{pq}_{st})$.

\begin{figure}[t]
        \includegraphics[trim=4cm 6.5cm 4cm 7cm, clip, width=.8\columnwidth]{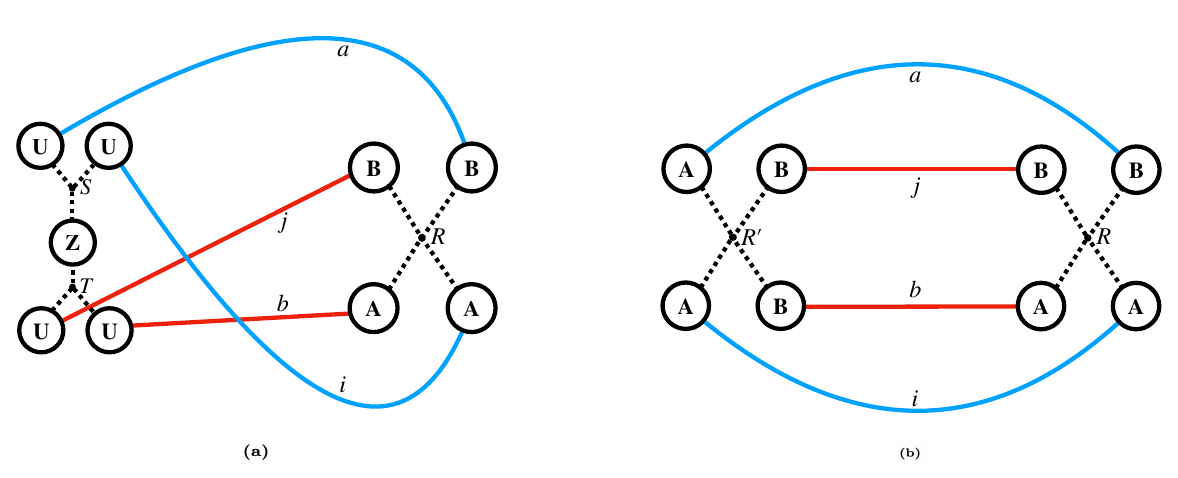}
    \caption{The LT MP2 correlation energy tensor-network. (a) utilizes a mixture of the THC approximation of the TEI tensor and the CPD approximation of the TEI tensor and (b) utilizes the CPD approximation for both tensors in the network.}
    \label{fig:cpd-contract}
\end{figure}

The CPD finds an algorithmic advantage over the THC because all factors of a CPD are connected by a single hyper-index, $R_\mathrm{CPD}$.
This is best demonstrated in \cref{fig:cpd-contract}, which shows CPD approximations of \cref{fig:thc-contract}b and \cref{fig:df-contract}a.
As we discuss in the next section, the computational complexity of contracting the networks in \cref{fig:cpd-contract} is $\mathcal{O}(N^3)$ assuming both $R_\mathrm{THC}$ and $R_\mathrm{CPD}$ are proportional to $N$, though this is an open question.
Although the CPD format is extremely powerful, its integration into electronic structure methods has been limited because its initialization is time-intensive. 
We introduce an optimization scheme that takes advantage of the THC tensor-network representation of the order-4 TEI tensor to more quickly optimize the an order-4 CPD of the TEI.

One most simple and effective ways to construct the CPD of the TEI tensor is via the analytic optimization of a CPD loss function
\begin{align}\label{eq:loss}
    f(\vec{x}) = \min_{\vec{x}} \frac{1}{2} \| g^{pq}_{st} - \tilde{g}^{pq}_{st}(\vec{x}) \|^2
\end{align}
using an alternating least squares (ALS)\cite{VRG:kroonenberg:1980:P,VRG:beylkin:2002:PNAS} optimization scheme.
Other effective schemes do exist like the tensor-cross interpolative-based regularized Newton method used previously by Benedikt et al.\cite{VRG:benedikt:2011:JCP,Benedikt:2013:MP,Benedikt:2013:JCP}
The CP-ALS optimization of \cref{eq:loss} has an iterative scaling of $\mathcal{O}(N^4R_\mathrm{CPD}) \approx \mathcal{O}(N^5)$. 
In our previous work\cite{VRG:pierce:2023:JCTC} we have shown that it is possible to approximate this CPD loss function by introducing the density-fitting approximation
\begin{align}\label{eq:df_loss}
    f(\vec{x}) \overset{\mathrm{DF}}{\approx} \min_{\vec{x}} \frac{1}{2} \| \sum_X B^{pX}_{s} B^{qX}_t - \tilde{g}^{pq}_{st}(\vec{x}) \|^2.
\end{align}
This approximation reduces the CPD optimization's computational complexity to $\mathcal{O}(N^4)$ and storage complexity to $\mathcal{O}(N^3)$ by allowing one to compute approximate gradients of \cref{eq:loss} in a matrix-free manner.
Here we study the approximation of \cref{eq:loss} using the THC approximation, instead of the DF approximation.
The THC-approximation CPD loss function can be written as 
\begin{align}\label{eq:lossF}
    f(\vec{x}) \overset{\mathrm{THC}}{\approx} \min_{\vec{x}} \frac{1}{2} \| \sum_{PQ} U^p_P V^P_s Z^P_Q U^q_Q V^Q_t - \tilde{g}^{pq}_{st}(\vec{x}) \|^2.
\end{align}
A full breakdown of the ALS optimization cost of \cref{eq:lossF} can be found in the supplemental information.
Using the THC approximated CPD loss-function reduces the complexity of computing approximate gradients of \cref{eq:loss} to $\mathcal{O}(NR^2)\approx \mathcal{O}(N^3)$ and reduces the storage complexity is $\mathcal{O}(R^2)\approx\mathcal{O}(N^2)$, where $R$ is either the CPD or THC rank.

Now we can leverage the efficient construction of the accurate THC approximation to rapidly optimize the order-4 CPD representation of the TEI.
With this CPD optimization method, we set the stage for the efficient construction of novel, reduced-scaling electronic structure method via the careful combination of the CPD and THC approximations.
Next, we will demonstrate the accuracy and efficiency of this combined THC and CPD approach via its integration into the LT approximated second order M{\o}ller-Plesset (MP2) method. 

\subsection{Reduced-Scaling CPD+THC LT MP2}
The spin-restricted, closed-shell MP2 perturbative correction to the Hartree-Fock (HF) energy is written as follows
\begin{align}\label{eq:Emp2}
    E_{\mathrm{MP2}} = \sum_{abij} \frac{g^{ab}_{ij} (2.0 g^{ab}_{ij} - g^{ba}_{ij})}{D^{ab}_{ij}}
\end{align}
where the HF energy denominator $D^{ab}_{ij} = \epsilon_a + \epsilon_b - \epsilon_i - \epsilon_j$ and $\epsilon$ is an eigenvalue of the diagonal Fock operator.
We use the indices $i,j,k,...$ to represent the set of occupied molecular orbitals and the indices $a,b,c,...$ to represent the set of unoccupied molecular orbitals.
In the following, we introduce the THC approximated LT MP2 algorithm, first derived by Martinez et al\cite{VRG:hohenstein:2012:JCP,VRG:kokkilaschumacher:2015:JCTC}, then we derive a low-scaling CPD approximated exchange component using the robust approximation.
This combined CPD+THC LT MP2 algorithm has a reduced computational scaling from $\mathcal{O}(N^2R_\mathrm{THC}^2)$ to $\mathcal{O}(NR_\mathrm{THC}R_\mathrm{CPD})$ and a reduced storage complexity from $\mathcal{O}(NR^2_\mathrm{THC})$ to $\mathcal{O}(R^2)$ where $R$ represents either $R_\mathrm{THC}$ or $R_\mathrm{CPD}$.\footnote{The scaling for the THC based LT MP2 algorithm derives from the scheme 2 figure in Ref. \citenum{VRG:kokkilaschumacher:2015:JCTC}}
We start our derivation by introducing the LT of the energy denominator tensor.

It is well understood that the energy denominator tensor can be represented as
\begin{align}\label{eq:lt_D}
    [D^{ab}_{ij}]^{-1} = \int_0^\infty e^{-D^{ab}_{ij}t} dt \approx \sum_\alpha w_\alpha e^{-D^{ab}_{ij}t_\alpha}
\end{align}
using the LT.\cite{VRG:almlof:1991:CPL,Haser:1998:JCP,VRG:haser:1993:TCA,VRG:wilson:1997:TCA}
As shown in \cref{eq:lt_D}, the exact integration over $t$ can be performed on a grid of $\alpha$ quadrature points.
It has been shown that the number of quadrature points is independent of system size and as few as 10 points can already produce MP2 energies with micro-hartree accuracy in some systems.\cite{Haser:JCP:1992}
For convenience, the LT can be factorized into one of the TEI tensors to form an approximation to the MP1 wavefunction tensor, $\mathcal{T}$,
\begin{align}\label{eq:mp1_t}
    t^{a_\alpha b_\alpha}_{i_\alpha j_\alpha} =\Gamma^{a}_{\alpha} \Lambda^\alpha_{i} g^{ab}_{ij}  \Gamma^{b}_\alpha  \Lambda^\alpha_{j} 
\end{align}
$\mathbf \Gamma$ and $\mathbf \Lambda$ are the LT transformation matrices
\begin{align}
    \Gamma^{a}_\alpha &= |w_\alpha|^\frac{1}{4} \exp((\epsilon_f - \epsilon_a) t_\alpha / 2) |a \rangle \\ \nonumber
    \Lambda^{\alpha}_i &= |w_\alpha|^\frac{1}{4} \exp((\epsilon_i - \epsilon_f) t_\alpha / 2) |i \rangle
\end{align}
and $\epsilon_f$ is a value that lies between the HOMO and LUMO of the molecule.
With the LT approximation, the MP2 energy can be expressed as
\begin{align}\label{eq:LT-Emp2}
    E_{\mathrm{MP2}} \overset{\mathrm{LT}}{\approx} \sum_\alpha \sum_{abij} t^{a_\alpha b_\alpha}_{i_\alpha j_\alpha} (2.0 g^{ab}_{ij} - g^{ba}_{ij})
\end{align}
It is important to notice that $\alpha$ is, in essence, the hyper-rank of the LT. Since there is no interaction between different values of $\alpha$, the computation is trivially parallelizable over $\alpha$. As a result, readers can essentially ignore the $\alpha$ in the summation over $a,b,i$ and $j$ and instead consider $\alpha$ as the number of times these summations must be performed.

Next, we introduce the THC approximation into the LT MP2 expression, \cref{eq:LT-Emp2}.
There are three components to the LT MP2 algorithm, the construction of the approximate MP1 wavefunction, the computation of the MP2 correlation energy term, and the computation of the MP2 exchange energy term.
First, one can see that the substitution of the THC approximation in the LT approximation, \cref{eq:mp1_t}, reduces the complexity of transformation from $N^4$ to $N^2$:
\begin{align}
    \bar{t}^{a_\alpha b_\alpha}_{i_\alpha j_\alpha} = \sum_{PQ} (\Gamma^{a}_{\alpha} U^a_P)  (\Lambda^\alpha_{i} V^P_i) Z^P_Q  (\Gamma^{b}_{\alpha} U^b_Q) (\Lambda^\alpha_{j} V^Q_j) \overset{\mathrm{ISDF-LT}}{\approx} t^{a_\alpha b_\alpha}_{i_\alpha j_\alpha}.
\end{align}

Next, we introduce the THC approximated MP2 correlation ($J$) term.
The THC approximated $J$ term can be written as the following
\begin{align}\label{eq:thc_Ej}
    E_\mathrm{MP2_J} &= 2 \sum_{abij} t^{ab}_{ij} g^{ab}_{ij} \\ \nonumber
    &\overset{\mathrm{ISDF-LT}}{\approx} 2 \sum_\alpha \bar{t}^{ab}_{ij} \bar{g}^{ab}_{ij}
    \\ \nonumber
    &= 2 \sum_\alpha 
    \sum_{PQST} (\sum_a U^{a_\alpha}_P U^a_S)  (\sum_i V^P_{i_\alpha} V^S_i)(\sum_b U^{b_\alpha}_Q U^b_T) (\sum_j V^Q_{j_\alpha} V^T_j) Z^P_Q Z^S_T \\ \nonumber
    &= 2 \sum_\alpha [\sum_{QT}  (\sum_{S}(\sum_{P} U^P_S V^P_S Z^P_Q) Z^S_T) U^Q_T V^Q_T]_\alpha
\end{align}
By contracting \cref{eq:thc_Ej} in the order specified by the parenthesis, one can see that the complexity of computing the $J$ term in this way is $\mathcal{O}(R_\mathrm{THC}^3)$.
We can also introduce the THC approximation into the LT MP2 exchange ($K$) term, i.e.
\begin{align}\label{eq:thc_Ek}
    E_\mathrm{MP2_K} &\overset{\mathrm{ISDF-LT}}{\approx} - \sum_\alpha \sum_{abij} \bar{t}^{a_\alpha b_\alpha}_{i_\alpha j_\alpha} \bar{g}^{ba}_{ij}. \\ \nonumber
    &= - \sum_{\alpha PQST} (\sum_{a}U^{a_\alpha}_P U^a_T)(\sum_{b} U^{b_\alpha}_{Q} U^b_S) (\sum_{i} U^P_{i_\alpha} U^S_i) (\sum_{j}U^Q_{j_\alpha}U^T_j) Z^P_Q Z^S_T \\ \nonumber
    &= -\sum_{\alpha PQST} (U^T_P U^S_Q Z^P_Q Z^S_TU^S_P U^T_Q)
\end{align}
As demonstrated by Hohenstein et al\cite{VRG:hohenstein:2012:JCP} the computational scaling of this term is $\mathcal{O}(N^2R_\mathrm{THC}^2) \approx \mathcal{O}(N^4)$.
To reduce the computational scaling of this exchange term, we will leverage the order-4 CPD approximation of the TEI tensor.

Here we introduce the reduced-scaling CPD+THC LT MP2 exchange term.
One can recognize that it is possible to apply the CPD once to \cref{eq:thc_Ek}, by replacing $\mathcal{G}$ or $\mathcal{T}$, or twice by replacing both of these tensors with their CPD apprioximation.
Replacing a single tensor, i.e. $\mathcal{G}$, with its CPD approximation gives the following equation
\begin{align}\label{eq:cpd_single_k}
     E_\mathrm{MP2_K} &\overset{\mathrm{CP+THC \text{ }LT}}{\approx} - \sum_\alpha \sum_{abij} \bar{t}^{a_\alpha b_\alpha}_{i_\alpha j_\alpha} \tilde{g}^{ba}_{ij} \\ \nonumber
    &= - \sum_{\alpha r}
    [\sum_{P} (\sum_a U^{a_\alpha}_P B^a_r)  (\sum_i V^P_{i_\alpha} A^r_i)][\sum_Q(\sum_b U^{b_\alpha}_Q A^b_r) (\sum_j V^Q_{j_\alpha} B^r_j) Z^P_Q]] \\ \nonumber
    &= -\sum_{\alpha r} [\sum_{Pr} B^P_r A^r_P[\sum_Q A^Q_r B^r_Q Z^P_Q]]
\end{align}
By computing the contractions in the order of the parenthesis, one is able to form $E_\mathrm{MP2_K}$ with a cost of $\mathcal{O}(R_\mathrm{THC}R^2_\mathrm{CPD}) \approx \mathcal{O}(N^3)$ cost.
Because of the symmetry in the expression, there is no difference in the approximation of either the bare TEI tensor or the TEI tensor in the approximate MP1 wavefunction.
This is best illustrated with the following expression
\begin{align}
    \sum_{abij} (\Gamma^{a}_{\alpha} \Lambda^\alpha_{i} \bar{g}^{ab}_{ij}  \Gamma^{b}_\alpha  \Lambda^\alpha_{j}) \tilde{g}^{ba} _{ij} = \sum_{abij} \tilde{g}^{ab}_{ij} (\Gamma^{a}_{\alpha} \Lambda^\alpha_{i} \bar{g}^{ba}_{ij}  \Gamma^{b}_\alpha  \Lambda^\alpha_{j})
\end{align}
By replacing both of TEI tensors in $K$ one finds the following expression
\begin{align}
    E_\mathrm{MP2_K} &\overset{\mathrm{CPD+THC\text{ }LT}}{\approx} - \sum_\alpha \sum_{abij} \tilde{t}^{a_\alpha b_\alpha}_{i_\alpha j_\alpha} \tilde{g}^{ba}_{ij} \\ \nonumber
    &= - \sum_{\alpha r r^\prime} (\sum_a A^{a_\alpha}_r B^a_{r^\prime})(\sum_b B^{b_\alpha}_r A^b_{r^\prime}) (\sum_{i} A_{i_\alpha}^{r}B_i^{r^\prime} ) (\sum_j B^r_{j_\alpha} A^{r^\prime}_{j}).
\end{align}
The double approximated CPD exchange term clearly has a computational cost of $\mathcal{O}(NR^2_\mathrm{CPD}) \approx O(N^3)$.
It is important to note that this formulation is exactly the derived expression from the study of CP approximated MP2 by Benedikt et al.\cite{VRG:benedikt:2011:JCP}
Both the single and double CPD approximated $K$ term generate, formally, $\mathcal{O}(N^3)$ algorithms, however both of these formulations are subject to a degree of error introduced by CPD approximation.

As a means to eliminate leading-order CPD error in $E_\mathrm{MP2_K}$ we utilize the robust approximation.\cite{VRG:dunlap:2000:PCCPP, VRG:pierce:2021:JCTC} 
The robust CPD+ISDF approximation\footnote{Note, when we refer to \cref{eq:rob_K}, we exclude any robust label because we do not implement or test a non-robust version.} of the $K$ term can be expressed efficiently as
\begin{align}\label{eq:rob_K}
    E_\mathrm{MP2_K} \overset{\mathrm{CPD-ISDF-LT}}{\approx} - \sum_\alpha [2.0 * \sum_{abij} \bar{t}^{a_\alpha b_\alpha }_{i_\alpha j_\alpha} \tilde{g}^{ba}_{ij} - \sum_{abij} \tilde{t}^{a_\alpha b_\alpha }_{i_\alpha j_\alpha} \tilde{g}^{ba}_{ij}].
\end{align}
As demonstrated, the computational scaling of each component of the robust CPD+THC approximation is $\mathcal{O}(N^3)$ and this approach has a storage complexity of $\mathcal{O}(R_\mathrm{THC}R_\mathrm{CPD}) \approx O(R_\mathrm{CPD}^2)\approx \mathcal{O}(N^2)$.
In the remaining, we attempt to verify that the CPD+THC LT MP2 method can accurately and efficiently compute the LT MP2 energy . 

\section{Computational Details}\label{sec:comp}

All MP2 calculations have been computed using a developmental version of the massively parallel quantum chemistry (MPQC) software package\cite{VRG:calvin:2020:CR} 
and computations were run on the Flatiron Institute's Scientific Computing Core's (SCC) Rusty cluster, using Rome nodes equipped with 2 AMD EPYC 7742 processors.
Each processor has 64 CPU cores and 512 GB of memory.
The MP2 calculation uses a Legendre polynomial scheme to determine quadrature points for the Laplace transform\cite{Braess:2005:IJNA} of the energy denominator and the number of quadrature points is fixed to 12.
In this work, we make no attempt to analyze the dependence of the MP2 energy with respect to the LT approximation, this has been the subject of several previous studies.
Practically, the construction of the LT approximation is uncorrelated to the construction of the THC and CPD approximations.

It is also important to note that we make no attempt benchmark the construction of the THC or remark on the convergence of the THC-based MP2 energy; this analysis can also be found in previous studies.
We hypothesize that the construction and accuracy of the THC approximated TEI tensor should have little effect on the performance of the CPD approximation.
Therefore, we seek to understand the convergence of the CPD approximation using a number of THC rank-$R_\mathrm{THC}$ approximations to verify this hypothesis.

The THC is constructed using the CPD approximation of the order-3 DF integral tensor to find THC factor matrices for the molecular orbital. To compute the THC core tensor the following least squares problem was solved
\begin{align}
    f(Z) = \min_{Z} \frac{1}{2}\| \sum_{X} B^{aX}_{i} B^{bX}_{j} - \sum_{PQ} U^a_P V^P_i Z^P_Q U^b_Q V^Q_j \|^2.
\end{align}
As pointed out previously, there are currently no $\mathcal{O}(N^3)$ ISDF algorithms for molecules but we look forward to ongoing research in this direction.
Furthermore,when an $\mathcal{O}(N^3)$ ISDF implementations is available, our combined CPD+THC approach can take advantage of the algorithm with practically no modifications.
The order-3 CPD was computed using an implementation in the C++ open-source Basic Tensor Algebra Subroutine (BTAS) library\cite{BTAS} and 
the novel order-4 CPD approximation of the THC approximated TEI tensor was implemented in both the BTAS and the C++ open-source TiledArray libraries.\cite{Calvin:TA}
All CPD implementations use a ALS optimization strategy
and both CPD algorithms use an initialization scheme where tensors are filled randomly with numbers between [-1,1] drawn from a uniform distribution.
The CPD was considered converged when the relative change in fit fell below $10^{-3}$, where fit is defined as
\begin{align}
    \mathrm{fit} = \frac{\|T - \tilde{T}\|}{\|T\|}.
\end{align}
We have shown previously that such a convergence is sufficient for accurate chemistry applications.\cite{VRG:pierce:2021:JCTC,VRG:pierce:2023:JCTC}
To accurately portray the cost of the CPD+THC LT MP2 approach we define the cost of this new method as the time to compute the CPD plus the time to compute the MP2 energy, i.e.
\begin{align}
    t_\mathrm{CPD-ISDF-LT} = t_\mathrm{CPD-ALS} + t_\mathrm{MP2}.
\end{align}
Because computing the optimal THC is not the focus of this work we do not include the time of its construction in any of our results.

\begin{figure}[!b]
    \begin{subfigure}{0.5\textwidth}
        \includegraphics[width=\columnwidth]{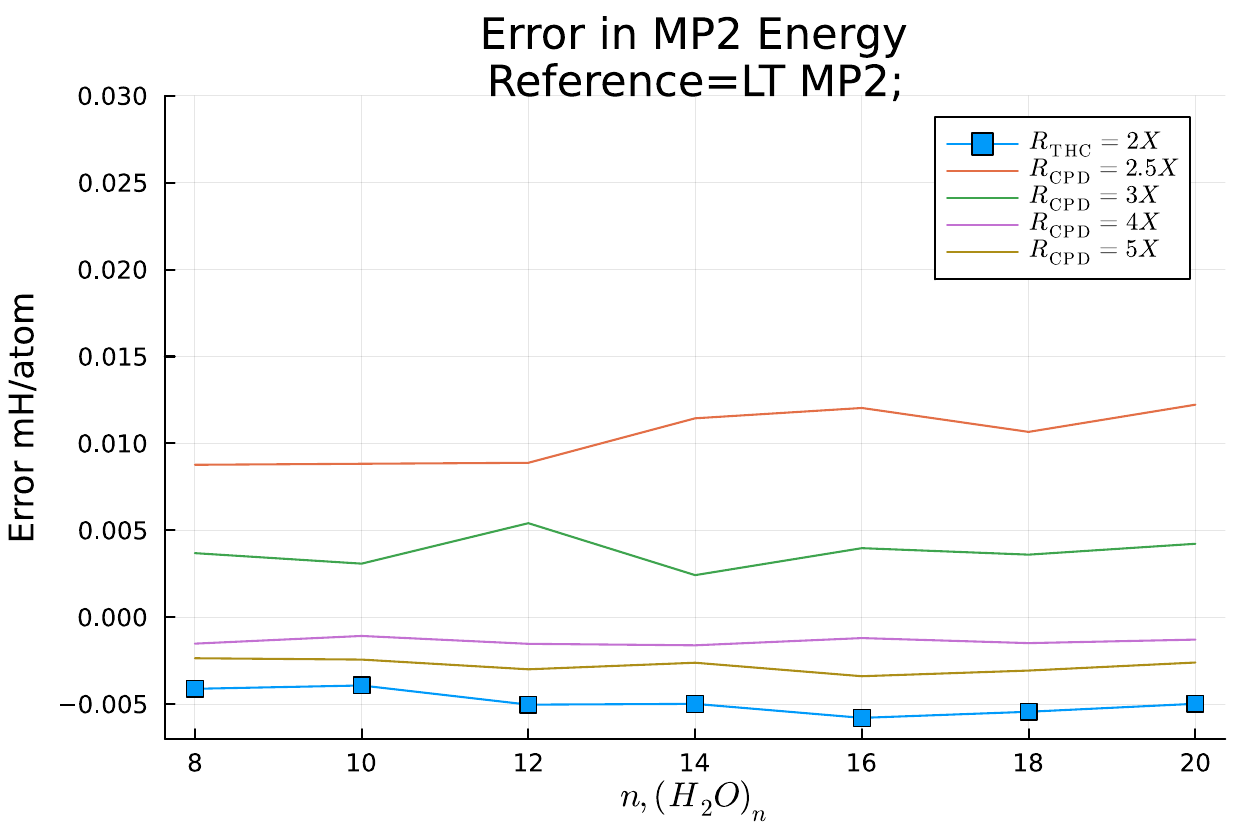}
        \caption{}
        \label{fig:1a}
    \end{subfigure}\hfill
    \begin{subfigure}{0.49\textwidth}
        \includegraphics[width=\linewidth]{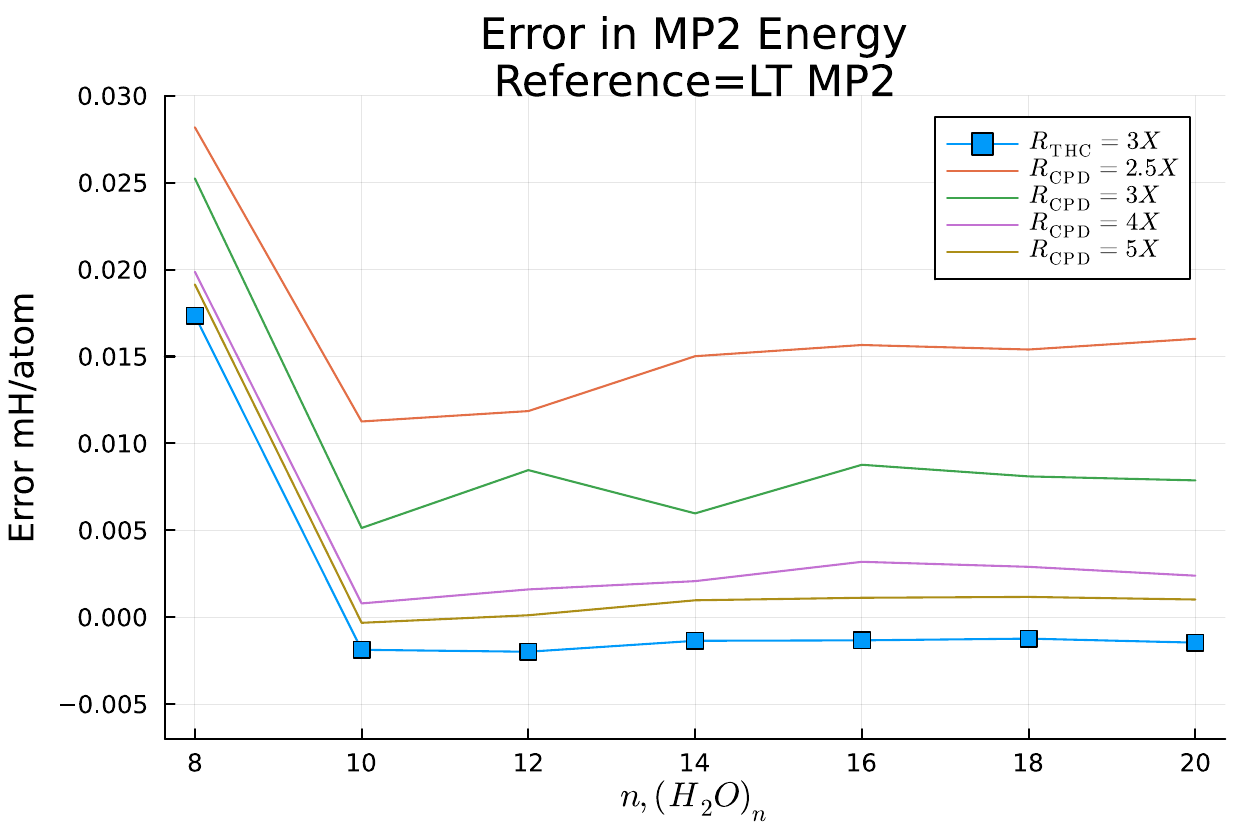}
        \caption{}
        \label{fig:1b}
    \end{subfigure}\hfill
    \begin{subfigure}{0.5\textwidth}
        \includegraphics[width=\columnwidth]{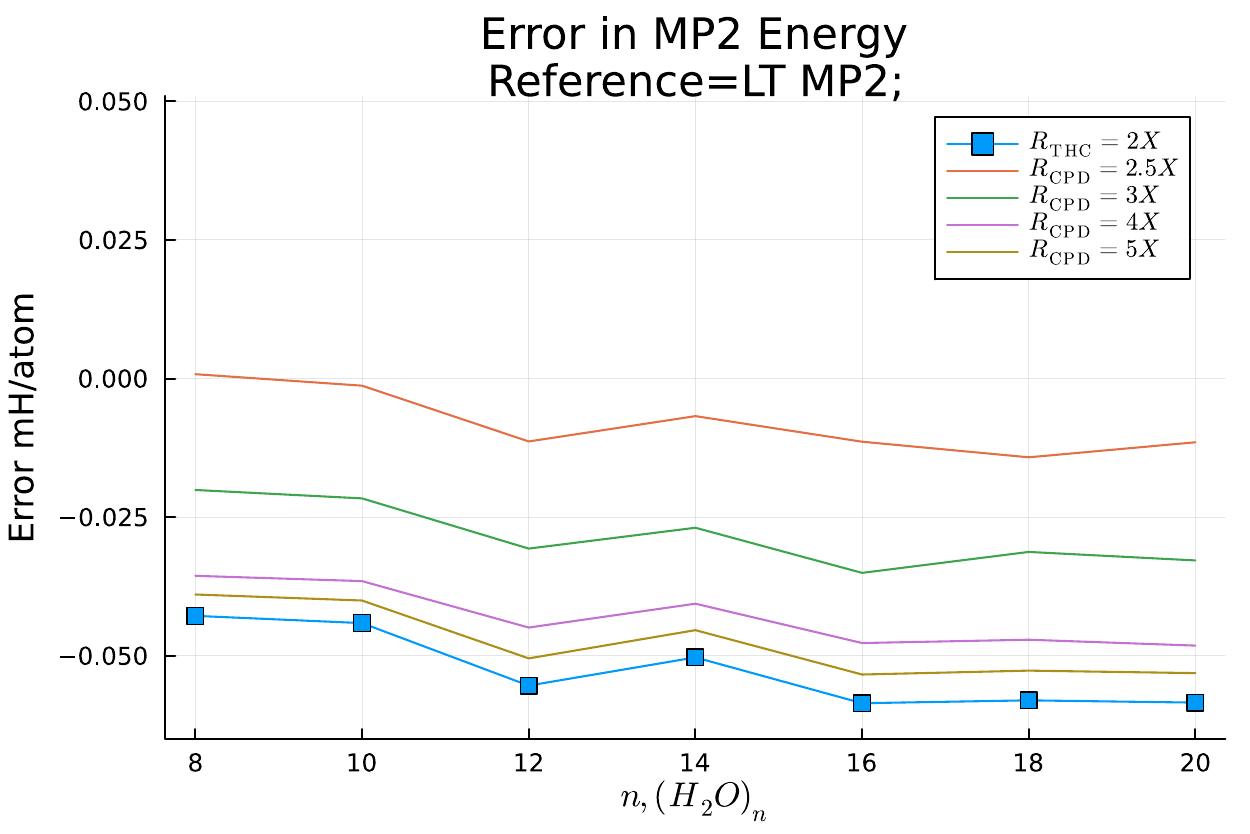}
        \caption{}
        \label{fig:1c}
    \end{subfigure}\hfill
    \begin{subfigure}{0.49\textwidth}
        \includegraphics[width=\linewidth]{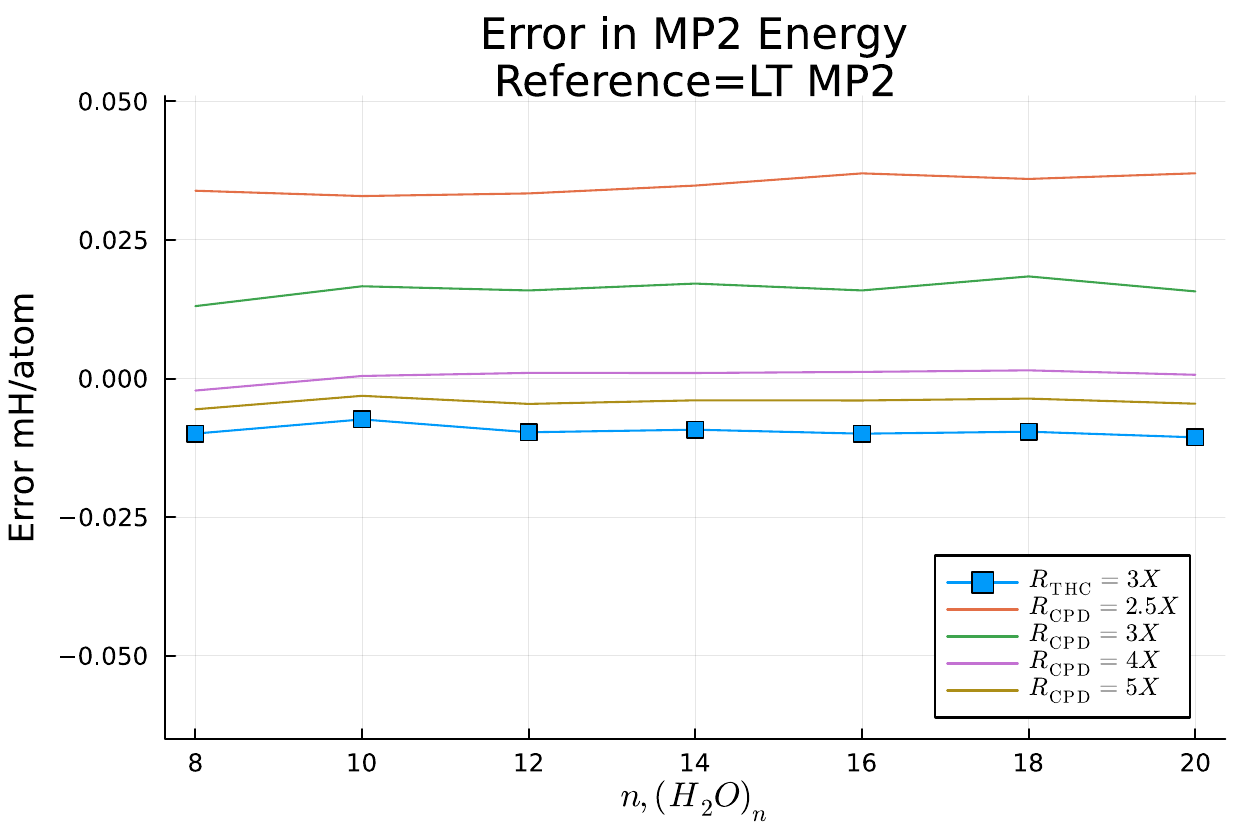}
        \caption{}
        \label{fig:1d}
    \end{subfigure}\hfill
\caption{MP2 error of the approximated LT MP2 methods for the set of water clusters with between 8 and 20 molecules. (a) and (b) use the DZ/DZ-RI basis and (c) and (d) uses the TZ/TZ-RI basis. The order-4 CPD was optimized using the TiledArray based ALS solver}\label{fig:all_v_lt_mp2_wat}
\end{figure}

The focus of our study is water clusters with between 1 to 20 molecules in the TIP4P optimized geometry\cite{jorgensen:1983:JCP,Wales:1998:CPL}. 
For this investigation we used two basis combinations: the cc-pVDZ (DZ) orbital basis set (OBS) accompanied with the corresponding cc-pVDZ-RI (DZ-RI)\cite{Dunning:1989:JCP,VRG:weigend:2002:JCP,Hattig:2005:PCCP}  density fitting basis set (DFBS); the cc-pVTZ (TZ)\cite{Dunning1989,Kendall1992} OBS accompanied with the corresponding cc-pVTZ-RI (TZ-RI)\cite{Weigend2002a} DFBS.
Additionally, we compute the energy of 6 of the 7 clusters in the L7 dataset\cite{Ballesteros:JCP:2021} in the TZ/TZ-RI OBS/DFBS.\footnote{The clusters ordered from fewest to largest number of non-hydrogen atoms is (1) Circumcoronene GC base pair, (2) Circumcoronene adenine, (3) Phenylalanine residues trimer (4) Guanine trimer (5) Octadecane dimer (6) Coronene dimer.}

\section{Results and Discussion}\label{sec:results}

In \cref{fig:all_v_lt_mp2_wat} we show the error of approximating the LT MP2 method using the THC LT MP2 and the CPD+THC LT MP2 methods. In \cref{fig:1a,fig:1b} we use the DZ/DZ-RI basis sets and in \cref{fig:1c,fig:1d} we use the TZ/TZ-RI basis sets also the THC LT MP2 method is represented with the square markers on the plot whereas the CPD+THC LT MP2 results use no marker.
In these plots we start our analysis at the 8 water cluster because of a convergence issue in our CP3 solver, related to the initial guess, though the results for all water clusters can be found in the provided supplemental information.
You can see a relic of this convergence issue in the accuracy of the 8 water cluster in \cref{fig:1b}.
It is well known that the quality of a CP-ALS optimization is initial guess dependent and therefore, these isolated extraneous results are not representative of the THC approximation's high accuracy and can be corrected by choosing a different initial guess or THC initialization strategy.

\begin{figure}[t!]
    \begin{subfigure}{0.5\textwidth}
        \includegraphics[width=\columnwidth]{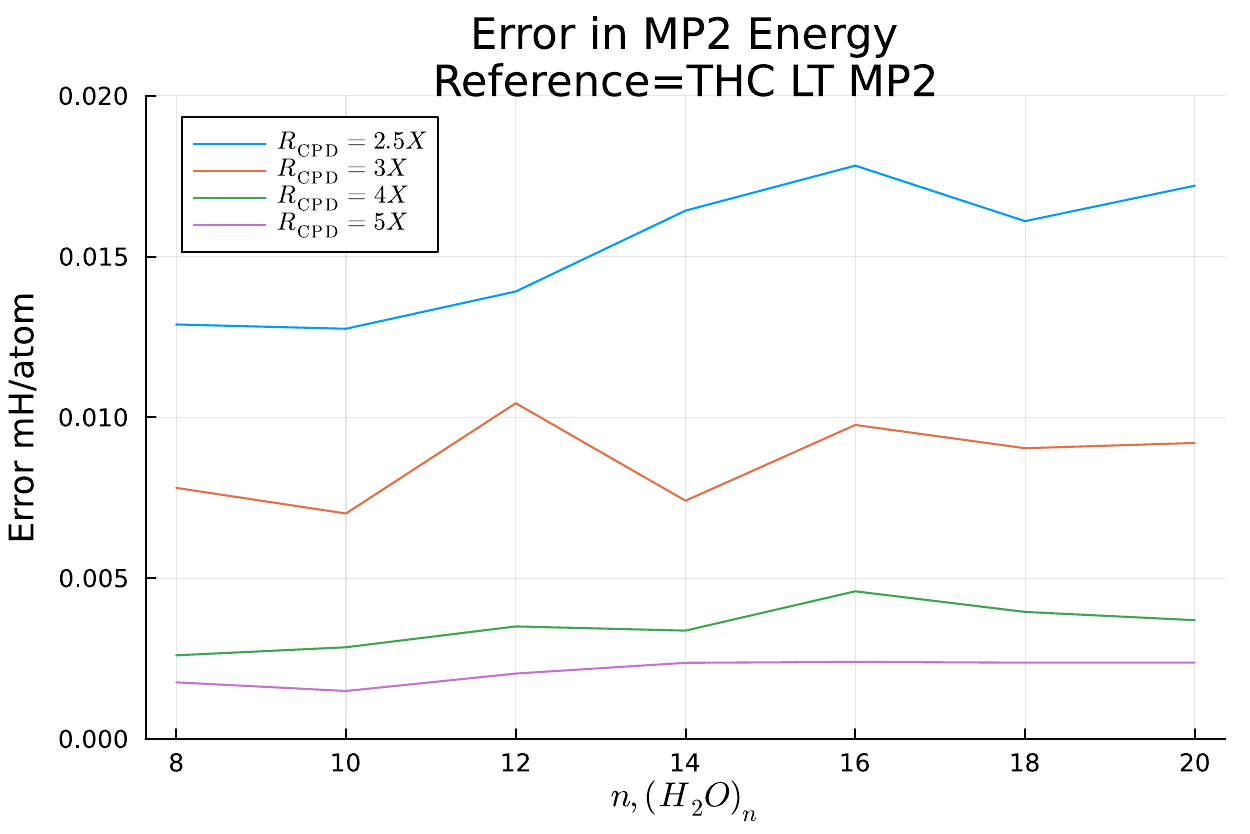}
        \caption{}
        \label{fig:2a}
    \end{subfigure}\hfill
    \begin{subfigure}{0.49\textwidth}
        \includegraphics[width=\linewidth]{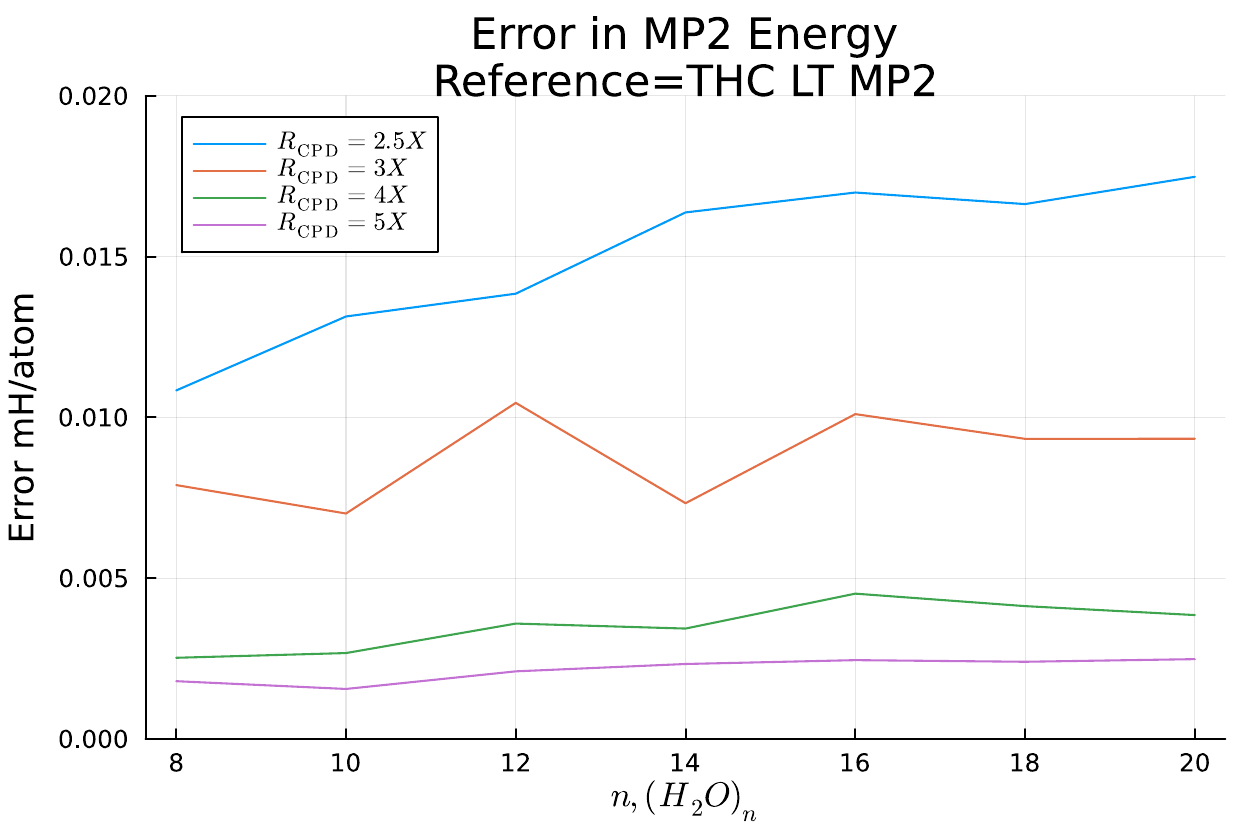}
        \caption{}
        \label{fig:2b}
    \end{subfigure}\hfill
    \begin{subfigure}{0.5\textwidth}
        \includegraphics[width=\columnwidth]{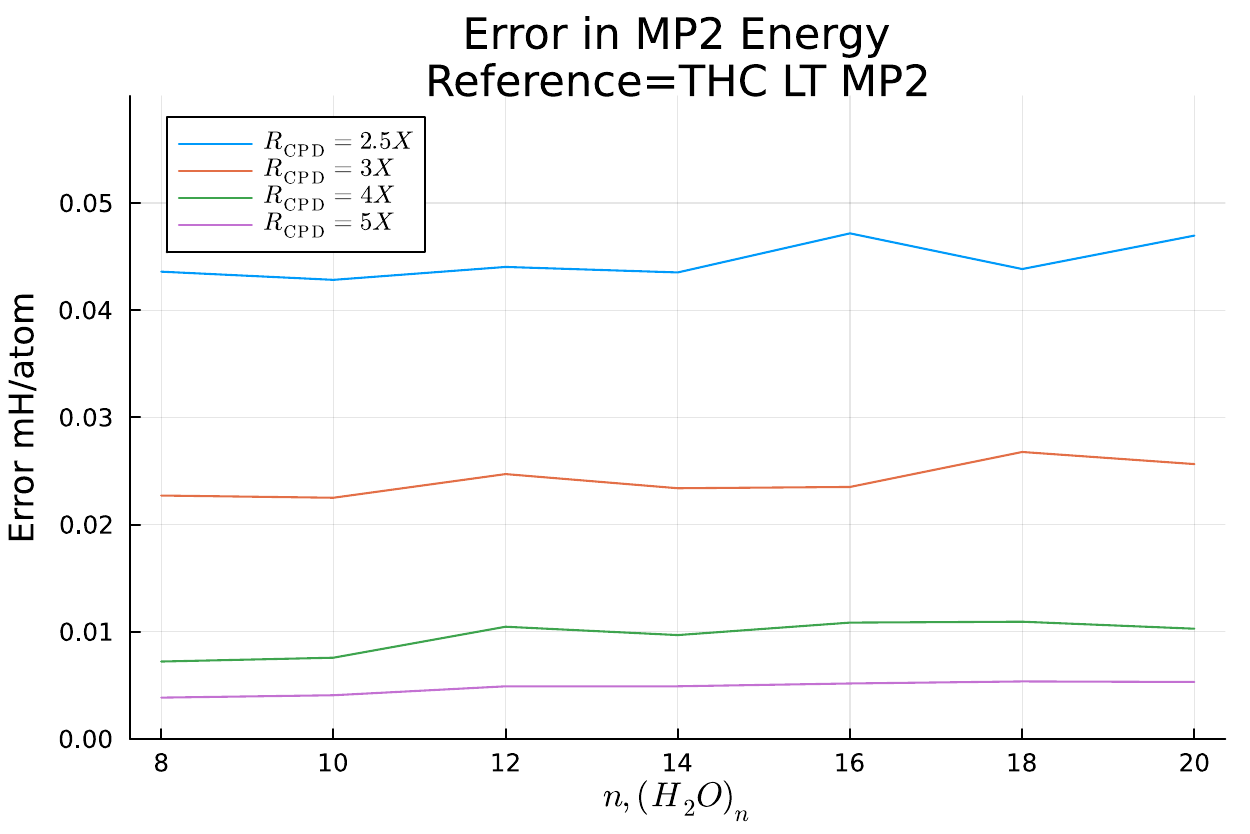}
        \caption{}
        \label{fig:2c}
    \end{subfigure}\hfill
    \begin{subfigure}{0.49\textwidth}
        \includegraphics[width=\linewidth]{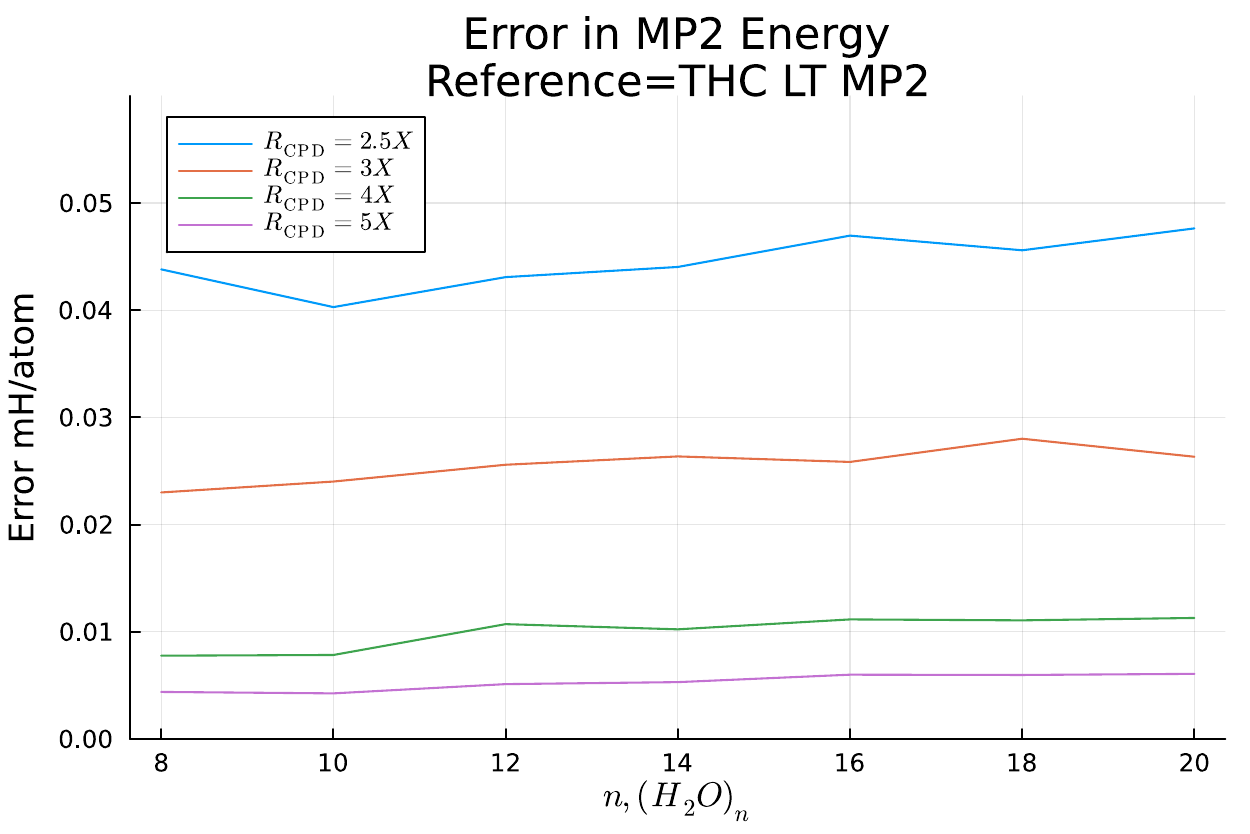}
        \caption{}
        \label{fig:2d}
    \end{subfigure}\hfill
\caption{Deviation in per non-hydrogen atom in the CP+THC LT MP2 energy from the THC LT MP2 energy for water clusters with between 8 and 20 water molecules. (a) uses the DZ/DZ-RI basis with a THC rank of $2X$, (b) uses the DZ/DZ-RI basis with a THC rank of $3X$,
(c) uses the TZ/TZ-RI basis with a THC rank of $2X$, and (d) uses the TZ/TZ-RI basis with a THC rank of $3X$. The order-4 CPD was optimized using the TiledArray based ALS solver}\label{fig:CP_v_isdf_mp2}
\end{figure}

From \cref{fig:all_v_lt_mp2_wat}, one can see that the accuracy of the CPD+THC LT MP2 is strongly dependent on the THC rank. 
In the analysis of absolute energies, we seek to introduce an error of less than $50$ $\mu$H per non-hydrogen atom with respect to the reference LT MP2 energy.
For the THC LT MP2 method, the DZ/DZ-RI basis requires a THC rank of just $2X$ to achieve this target accuracy whereas, the TZ/TZ-RI basis requires a slightly larger basis of $3X$.
These results align well with previous studies into the THC approximated LT MP2 method.\cite{VRG:hohenstein:2012:JCP,VRG:kokkilaschumacher:2015:JCTC,Lee:2019:JCTC}
The CPD+THC LT MP2 method appears to systematically underestimate the THC LT MP2 energy which can promote a favorable cancellation of error.
Because the errors associated with the CPD are relatively small, for molecules in the DZ/DZ-RI basis all studied values of $R_\mathrm{CPD}$ allows the CPD+THC LT MP2 to achieve the accuracy tolerance.
For the TZ/TZ-RI basis we see that when $R_\mathrm{THC} =2X$, values of $R_\mathrm{CPD}$ between $[2X-4X)$ achieve the accuracy tolerance and when $R_\mathrm{THC} =3X$, all studied values of $R_\mathrm{CPD}$ are withing the accuracy threshold.

In \cref{fig:CP_v_isdf_mp2} we look at the energy difference between the CPD+THC LT MP2 and the THC LT MP2 energy for the water clusters with between 8 and 20 molecules.
These figures are in the same order as \cref{fig:all_v_lt_mp2_wat} with DZ/DZ-RI in the top row,  TZ/TZ-RI in the bottom row.
Also, \cref{fig:2a,fig:2c} use a THC rank of $2X$ and \cref{fig:2b,fig:2d} use a THC rank of $3X$.
Using \cref{fig:CP_v_isdf_mp2}, we effectively isolate the difference in energy between the robust CPD approximated exchange term and the THC approximated exchange term.
One can see that increasing the CPD rank systematically eliminates the error in the CPD approximated exchange.
We recognize that the THC rank has little influence on the convergence of the CPD approximation.
The error of the CPD+THC LT MP2 appears to increase slowly with number of atoms.
However, when we compare the accuracy of the CPD+THC LT MP2 to the canonical LT MP2 method (i.e. \cref{fig:all_v_lt_mp2_wat}) we do not see this feature.
This could, again, be related to a fortuitous cancellation of errors as THC LT MP2 has been reported to slowly increase in error with molecule size.\cite{VRG:hohenstein:2012:JCP,VRG:kokkilaschumacher:2015:JCTC}

\begin{figure}[!t]
    \begin{subfigure}{0.5\textwidth}
        \includegraphics[width=\columnwidth]{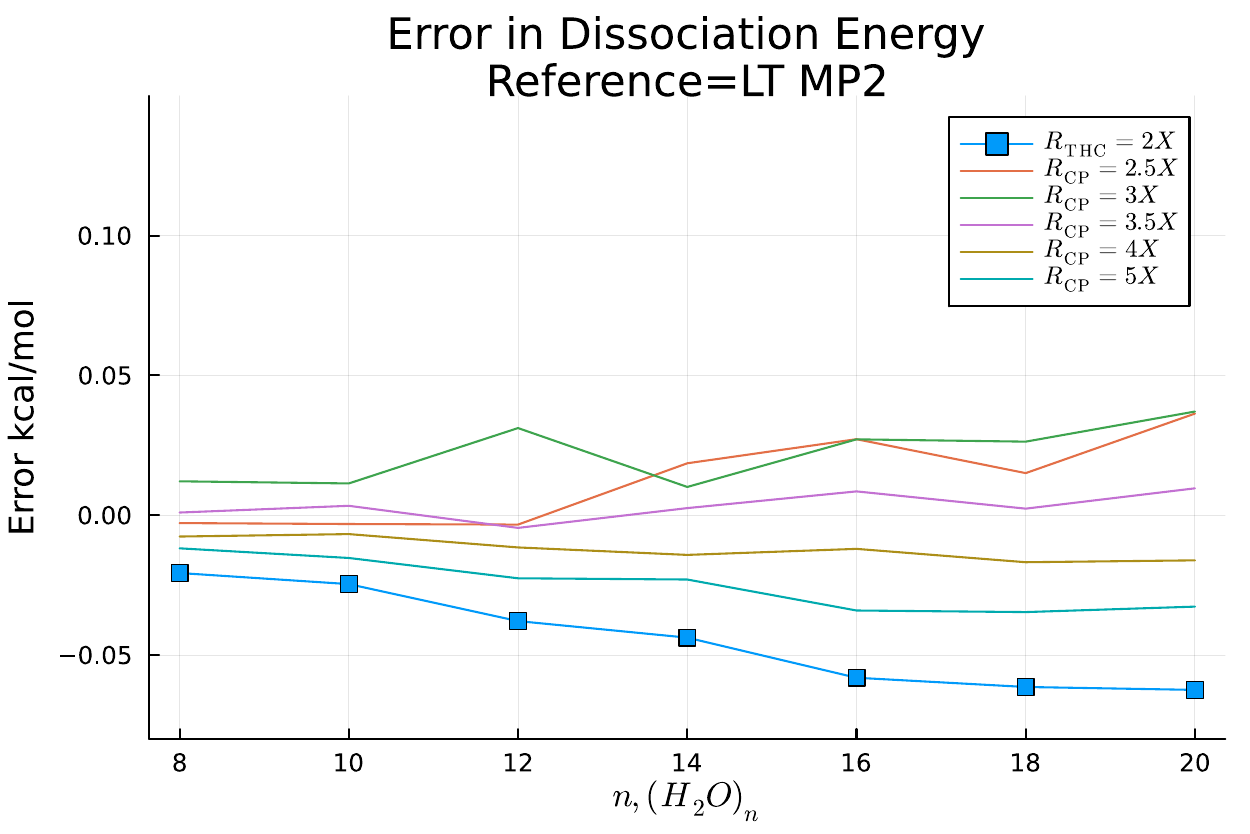}
        \caption{}\label{fig:3a}
    \end{subfigure}\hfill
    \begin{subfigure}{0.49\textwidth}
        \includegraphics[width=\linewidth]{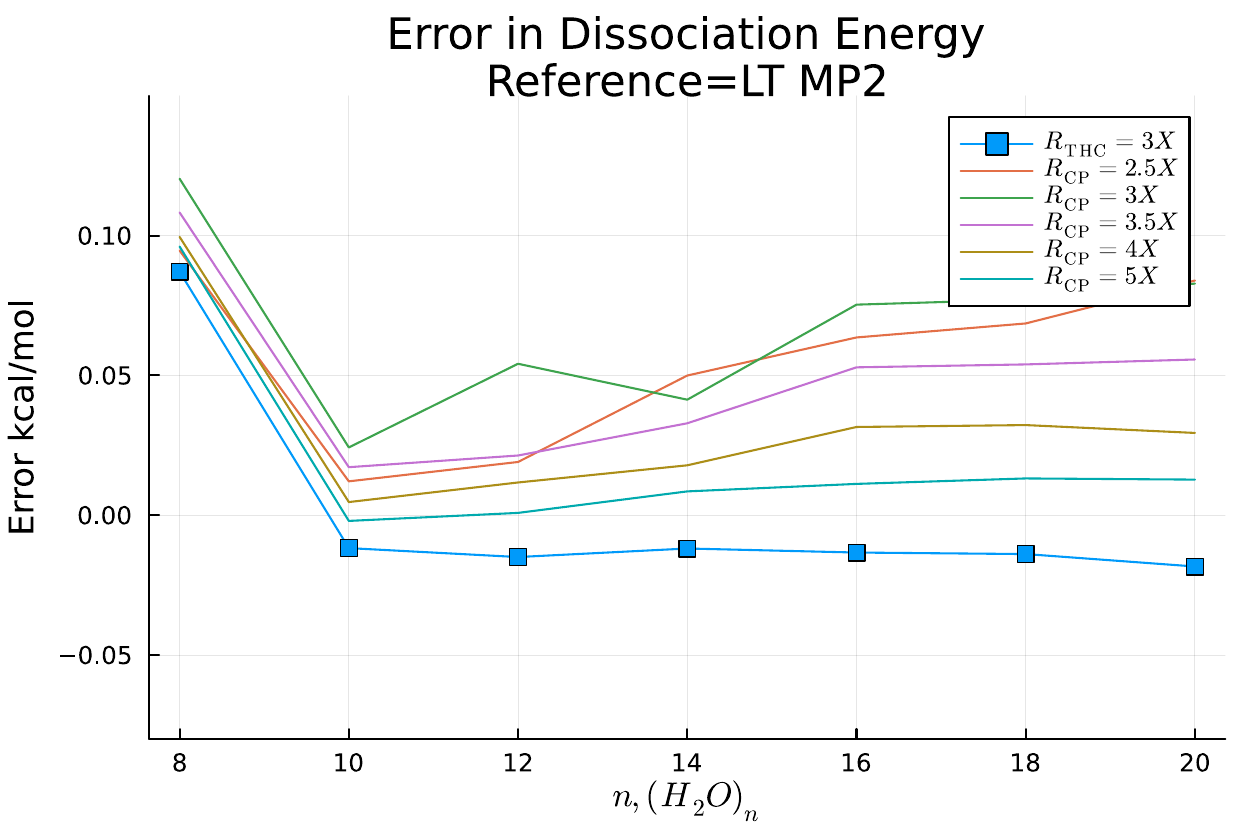}
        \caption{}\label{fig:3b}
    \end{subfigure}\hfill
    \begin{subfigure}{0.5\textwidth}
        \includegraphics[width=\columnwidth]{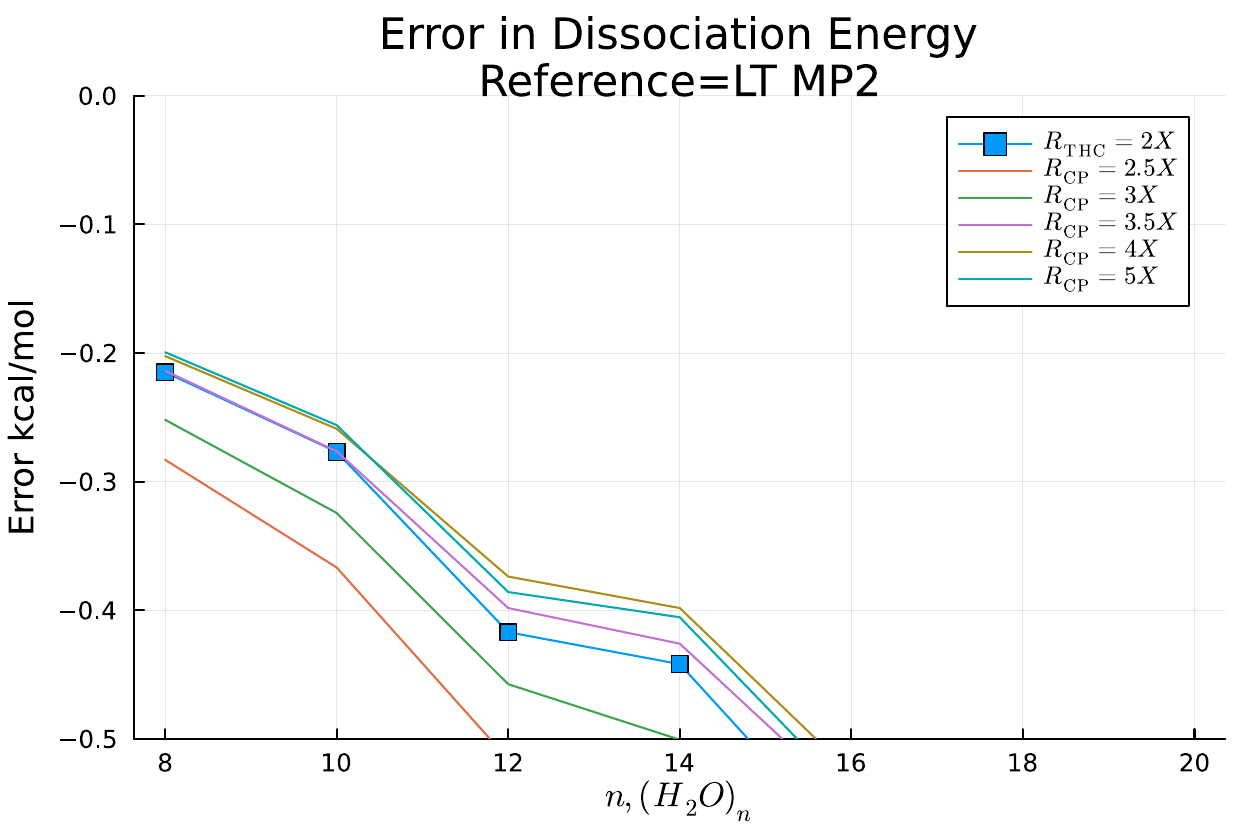}
        \caption{}\label{fig:3c}
    \end{subfigure}\hfill
    \begin{subfigure}{0.49\textwidth}
        \includegraphics[width=\linewidth]{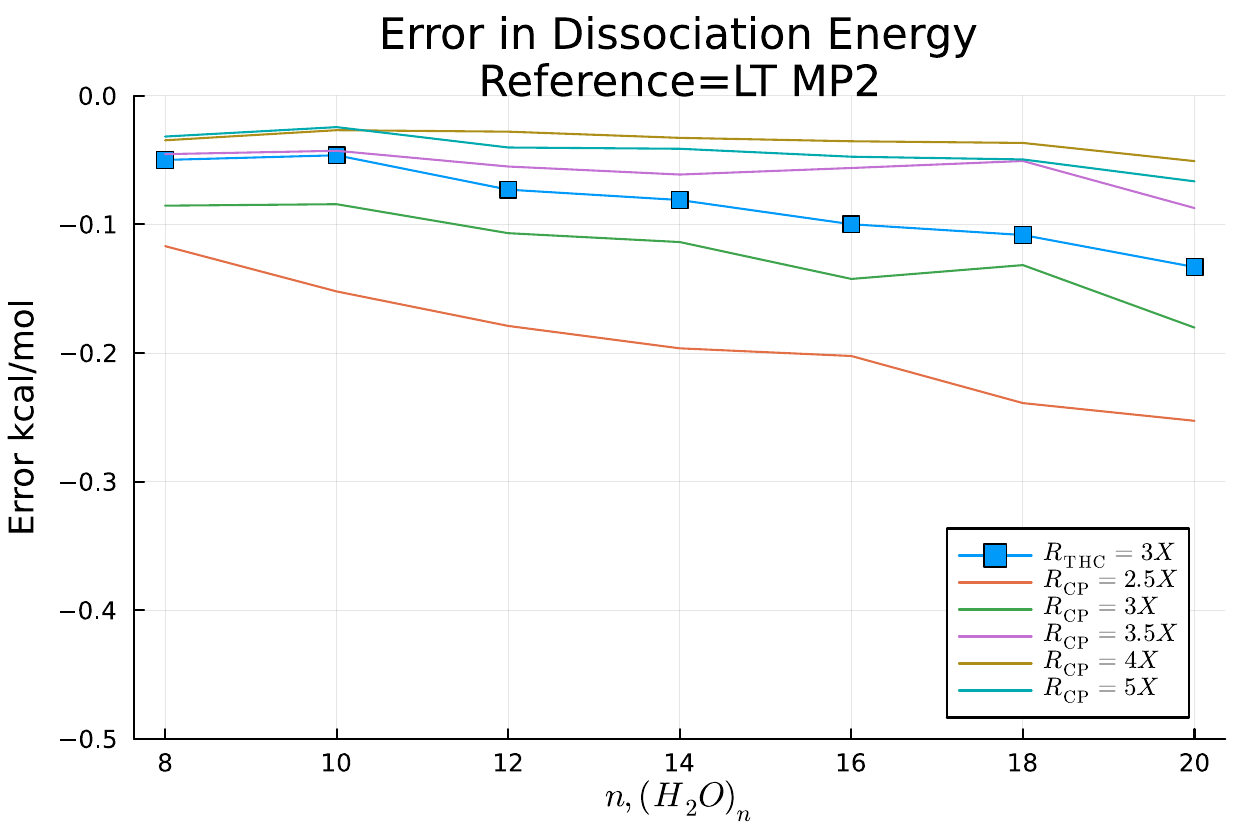}
        \caption{}\label{fig:3d}
    \end{subfigure}\hfill
\caption{Error in the MP2 dissociation energy of the approximated LT MP2 methods for water clusters with between 8 and 20 molecules. (a) and (b) uses the DZ/DZ-RI basis and (c) and (d) uses the TZ/TZ-RI basis. The order-4 CPD was optimized using the TiledArray based ALS solver}\label{fig:diss_error}
\end{figure}

In \cref{fig:diss_error} we present the error in the dissociation energy for the water clusters with between 8 and 20 molecules compared to the LT MP2 method. 
The order of these subfigures is equivalent to the order of \cref{fig:all_v_lt_mp2_wat,fig:CP_v_isdf_mp2}.
From \cref{fig:diss_error} we see that the accuracy of the CPD+THC LT MP2 dissociation energy is strongly dependent on the THC rank, consistent with the results from \cref{fig:all_v_lt_mp2_wat}.
For energy differences, we assess our data with a strict error tolerance of $0.1$ kcal/mol.
For the DZ/DZ-RI basis, we see that all of the studied values of the CPD+THC LT MP2 and THC LT MP2 method achieves the $0.1$ kcal/mol tolerance. 
It is interesting to notice that for small values of $R_\mathrm{CPD}$ using the DZ/DZ-RI basis the CPD+THC LT MP2 dissociation energies appear to increase with increasing molecule size.
This could be explained by the following, the relatively small THC approximation error is being overtaken by the CPD error at smaller values of $R_\mathrm{CPD}$ and, as seen in \cref{fig:CP_v_isdf_mp2}, this CPD error grows with molecular system size.
This artifact is reduced by increasing the CPD rank which is consistent with the observed reduction in the error associated with the CPD.

In the TZ/TZ-RI basis, similar to results in \cref{fig:all_v_lt_mp2_wat}, we find that the THC LT MP2 requires a larger value of $R_\mathrm{THC}$ compared to the DZ/DZ-RI basis.
And though the CPD+THC LT MP2 method does have a slight error cancellation effect on the computed energy differences, it is not able to significantly correct the errors found in \cref{fig:3c}.
For $R_\mathrm{THC}=3X$, \cref{fig:3d}, we see that THC LT MP2 method is able to achieve the target accuracy for small molecules but the dissociation energy error grows with increasing molecule size.
However, through fortuitous cancellation of errors, the CPD+THC LT MP2 method is able to correct the dissociation energy and accurately predict the value for all molecules for 
sufficiently large values of CPD rank, $R_\mathrm{CPD} \in (3X, 5X]$.
Increasing the CPD rank gradually converges to the THC LT MP2 results and, unlike the DZ/DZ-RI basis the CPD+THC LT MP2 does not over-correct the THC LT MP2 results.
From the results thus far, we can recommend a values of $R_\mathrm{THC} \geq 3 X$ and a $R_\mathrm{CPD} \geq 3X$ for the CPD+THC LT MP2 method.
For reference, we have included results for $R_\mathrm{THC}=4X$ in the supplemental information.
For this THC rank we see that, for all correctly converged THC optimizations, the THC LT MP2 method meets the prescribed absolute and dissociation energy accuracy tolerances and the CPD+THC LT MP2 method meets the accuracy tolerances for CP ranks values of $R_\mathrm{CPD} > 2.5X$.
Next, we investigate the wall-time performance of the CPD+THC LT MP2 method.

\begin{figure}[t]
    \begin{subfigure}{0.49\textwidth}
        \includegraphics[width=\linewidth]{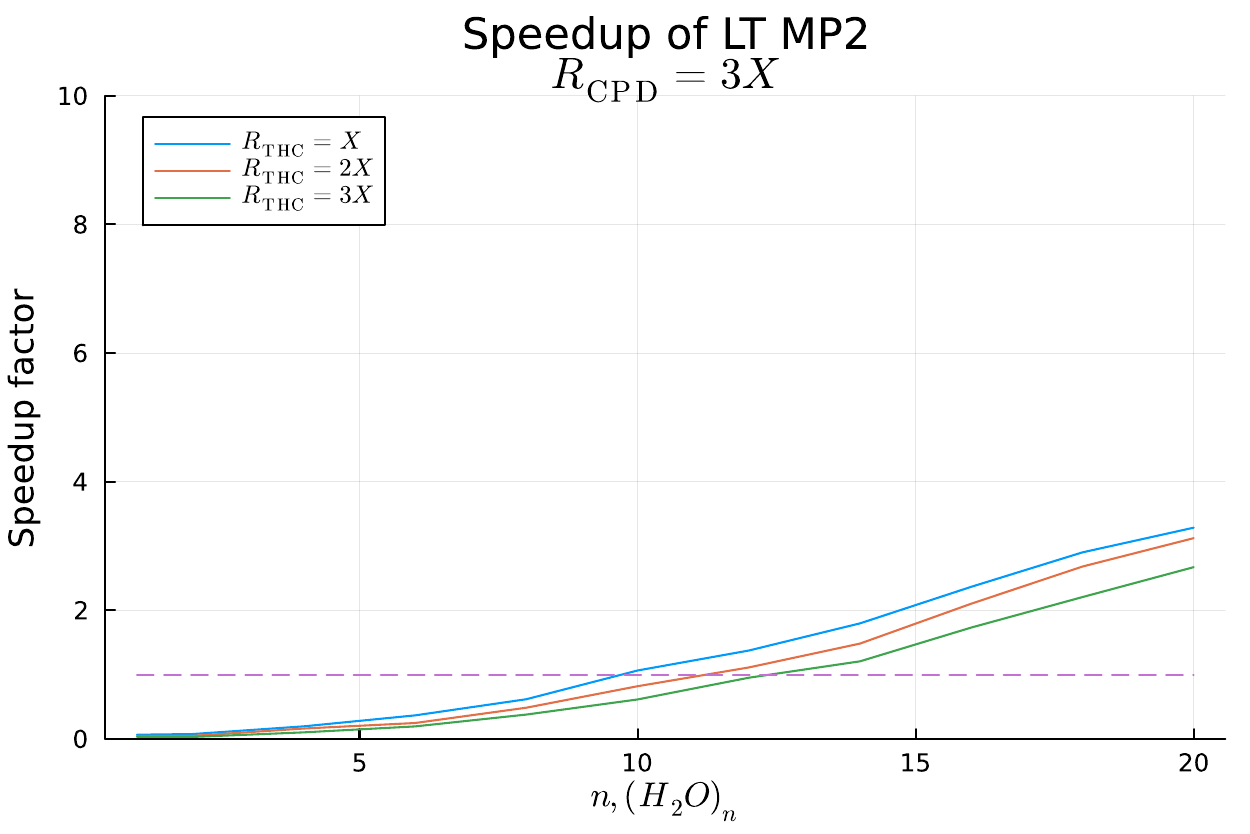}
        \caption{}\label{fig:4a}
    \end{subfigure}\hfill
    \begin{subfigure}{0.5\textwidth}
        \includegraphics[width=\columnwidth]{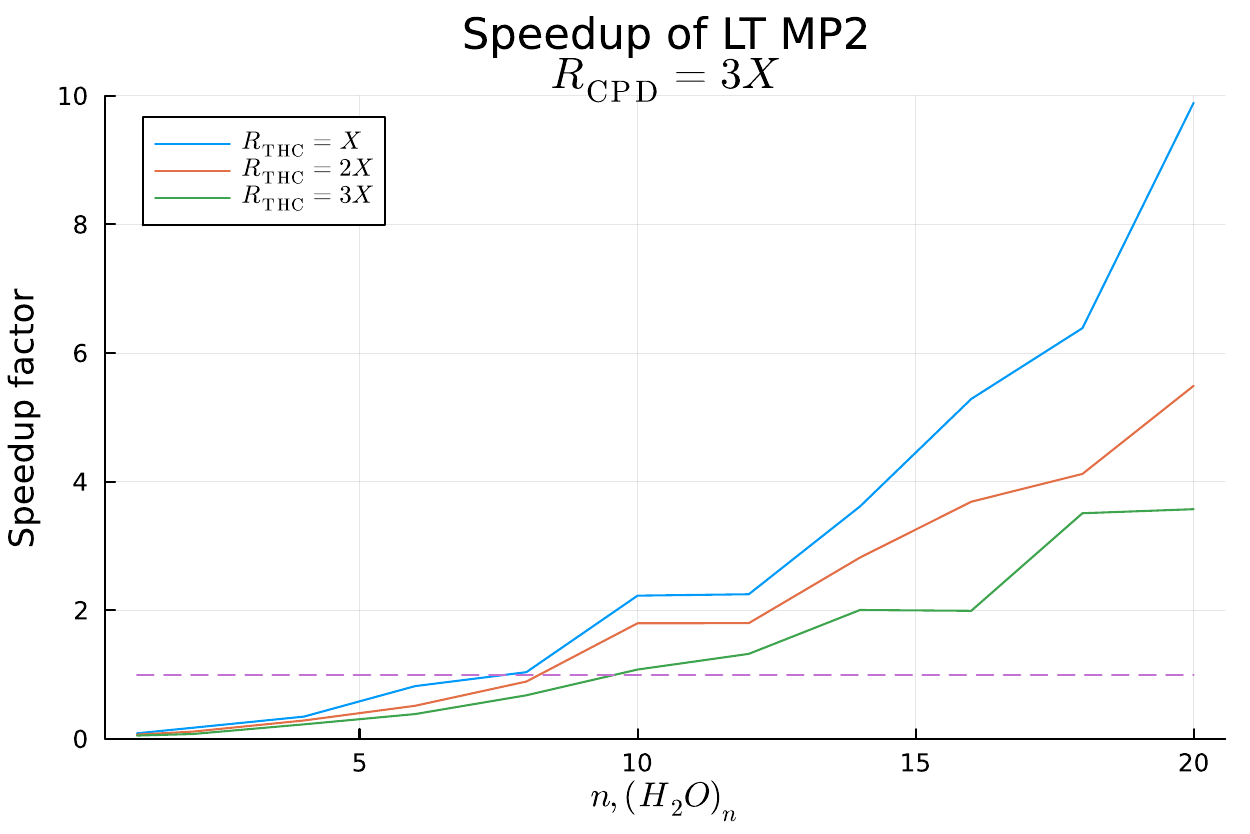}
        \caption{}\label{fig:4b}
    \end{subfigure}\hfill
    \begin{subfigure}{0.49\textwidth}
        \includegraphics[width=\linewidth]{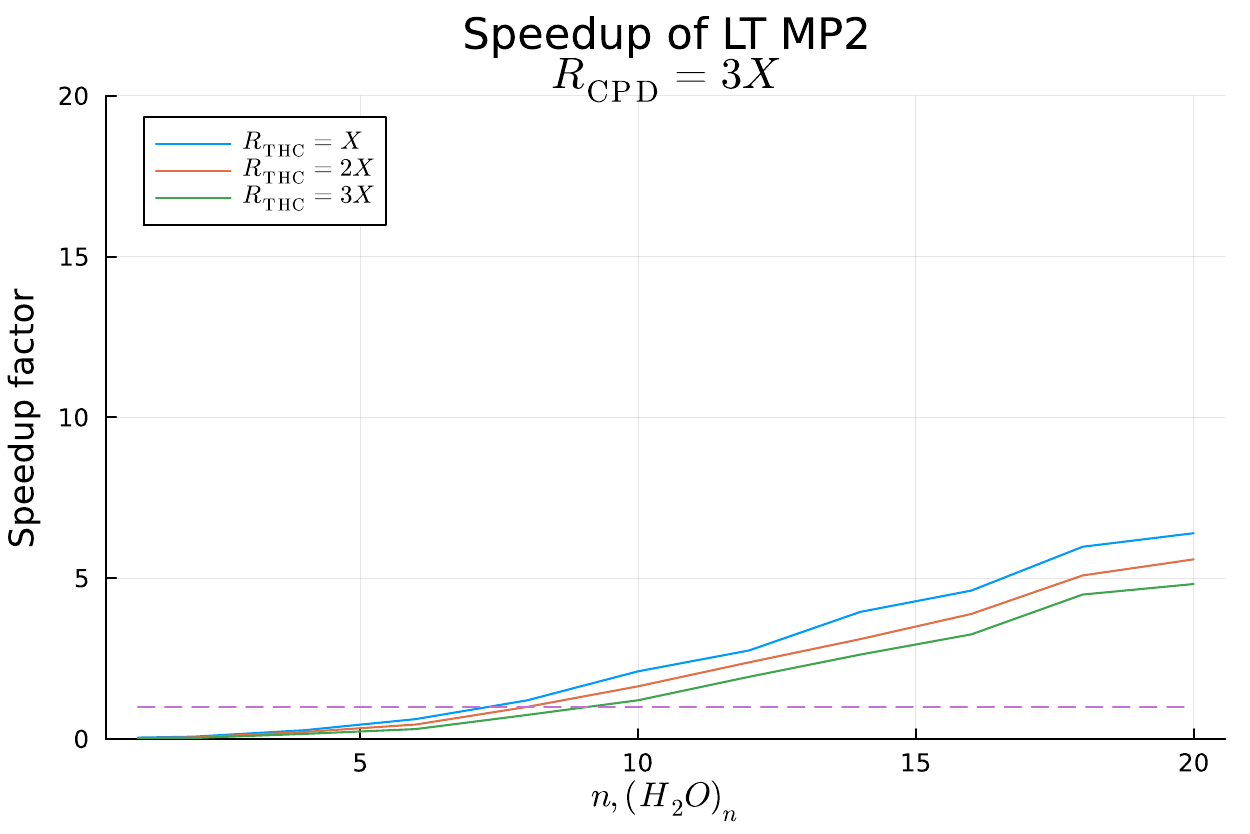}
        \caption{}\label{fig:4c}
    \end{subfigure}\hfill
    \begin{subfigure}{0.49\textwidth}
        \includegraphics[width=\linewidth]{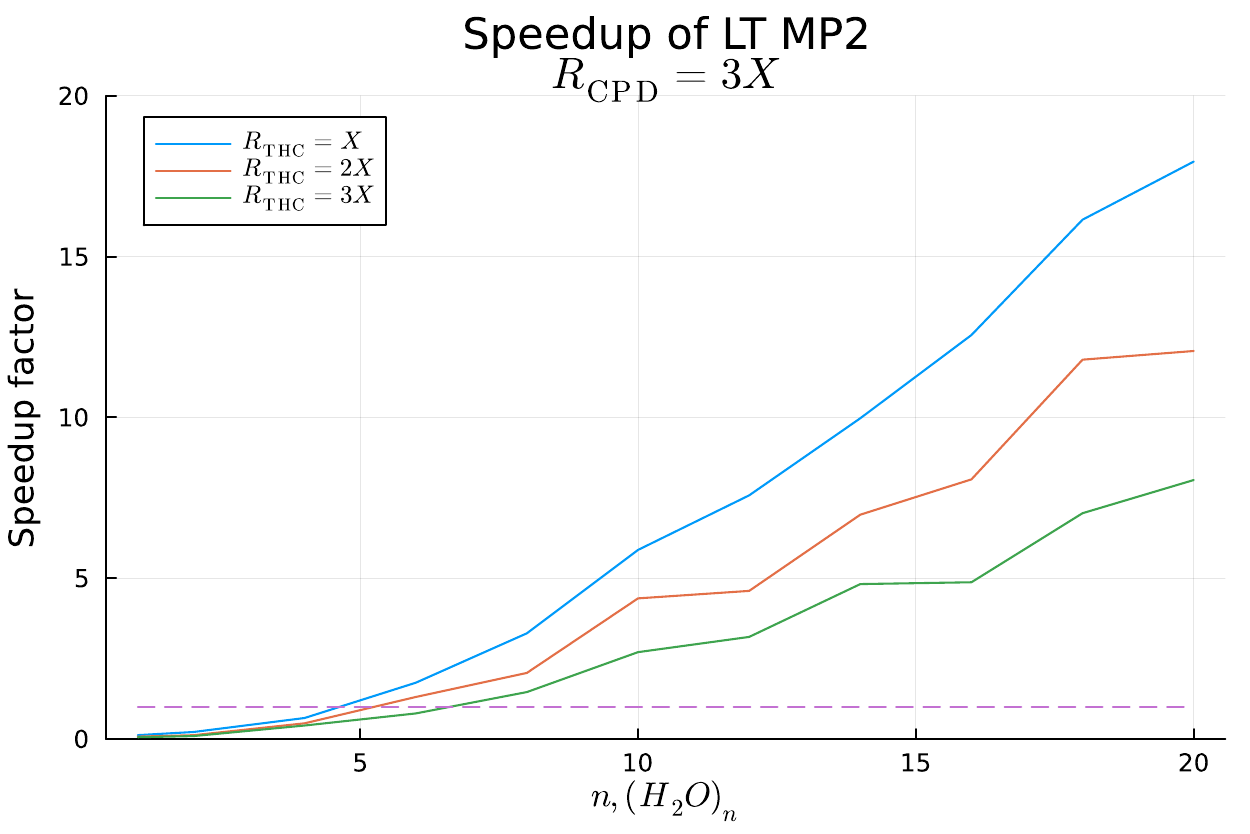}
        \caption{}\label{fig:4d}
    \end{subfigure}
\caption{Speedup of the CPD+THC LT MP2 method over the canonical LT MP2 method for the water clusters with between 1 and 20 molecules. Figures (a) and (b) use the DZ/DZ-RI basis and a THC rank of $3X$. Figures (c) and (d) use the TZ/TZ-RI basis and a THC rank of $3X$, respectively.
Figures (a) and (c) use a single Rome node with 128 CPU cores while figures (b) and (d) use a 16 CPU core partition of a Rome node. Also figures (a) and (c) use a TiledArray CP-ALS optimization algorithm while figures (b) and (d) use a BTAS CP-ALS optimization algorithm.}\label{fig:speedup_waters}
\end{figure}

In \cref{fig:speedup_waters} we illustrate the wall-time speedup of the CPD+THC LT MP2 over the LT MP2 method for water clusters with between 1 and 20 molecules using a CPD rank of $3X$.
\cref{fig:4a,fig:4b} studies the DZ/DZ-RI basis while \cref{fig:4c,fig:4d} studies the TZ/TZ-RI basis.
Please notice, the computations in \cref{fig:4a,fig:4c} were performed on a single Rome node with 128 CPU cores while the computations in \cref{fig:4b,fig:4d} were performed on a 16 CPU core partition of the node.
These partitioned cores were bound to a 128 GB NUMA core to eliminate intranode latency.
In \cref{fig:speedup_waters}, the dashed lines represents the cross-over for when the CPD+THC LT method is faster than the LT MP2 method. 

The choice of THC and CPD rank both impact the performance of the CPD+THC LT MP2 method.
We find that THC rank has a larger impact on the overall performance of the code.
Please note, speedup data for $R_\mathrm{CPD}=2X$ can be found in the supplemental information.
When using all 128 cores and the suggested $R_\mathrm{THC}$ of $3X$ and $R_\mathrm{CPD}$ of $3X$, \cref{fig:4a,fig:4c}, depending on the basis we find a performance benefit starting around 9-12 water molecule clusters (between 300-500 orbitals).

However, we noticed that our CPD optimization and MP2 method found a poor computational efficiency when tasked to run on all 128 CPU cores.
This is because our low-rank algorithms significantly reduces memory requirements and, therefore, these systems are much too small.
We can significantly improve the performance of our methods by using fewer computer resources and, because our methods have a storage complexity of just $\mathcal{O}(R^2)$, this is feasible for these test systems.
In \cref{fig:4b,fig:4d} we present the speedup data for the same computations run on a 16 CPU core partition of our 128 core Rome node.
We notice a significant improvement of performance across all calculations and, now, with the suggested $R_\mathrm{THC}$ of $3X$ and $R_\mathrm{CPD}$ of $3X$ we see a cross-over between 6 and 9 water molecule clusters (between 200-300 orbitals). 
Because the CPD+THC LT MP2 method has such a reduced memory overhead, the method would require significantly larger chemical systems to fully utilize the Rome node.
For example a THC and CP rank of 200,000 would require 320 GB of storage for the largest tensor intermediates which would map to a chemical problem with approximately 22,000 basis functions; a problem approximately 20 times larger than the 20 water molecule cluster.
Also, the CPD+THC LT MP2 method can be made more efficient by parallelizing the over quadrature points or taking advantage of GPU accelerators.

\begin{figure}[t]
        \includegraphics[trim=2cm 3cm 3cm 3cm, clip, width=0.8\textwidth]{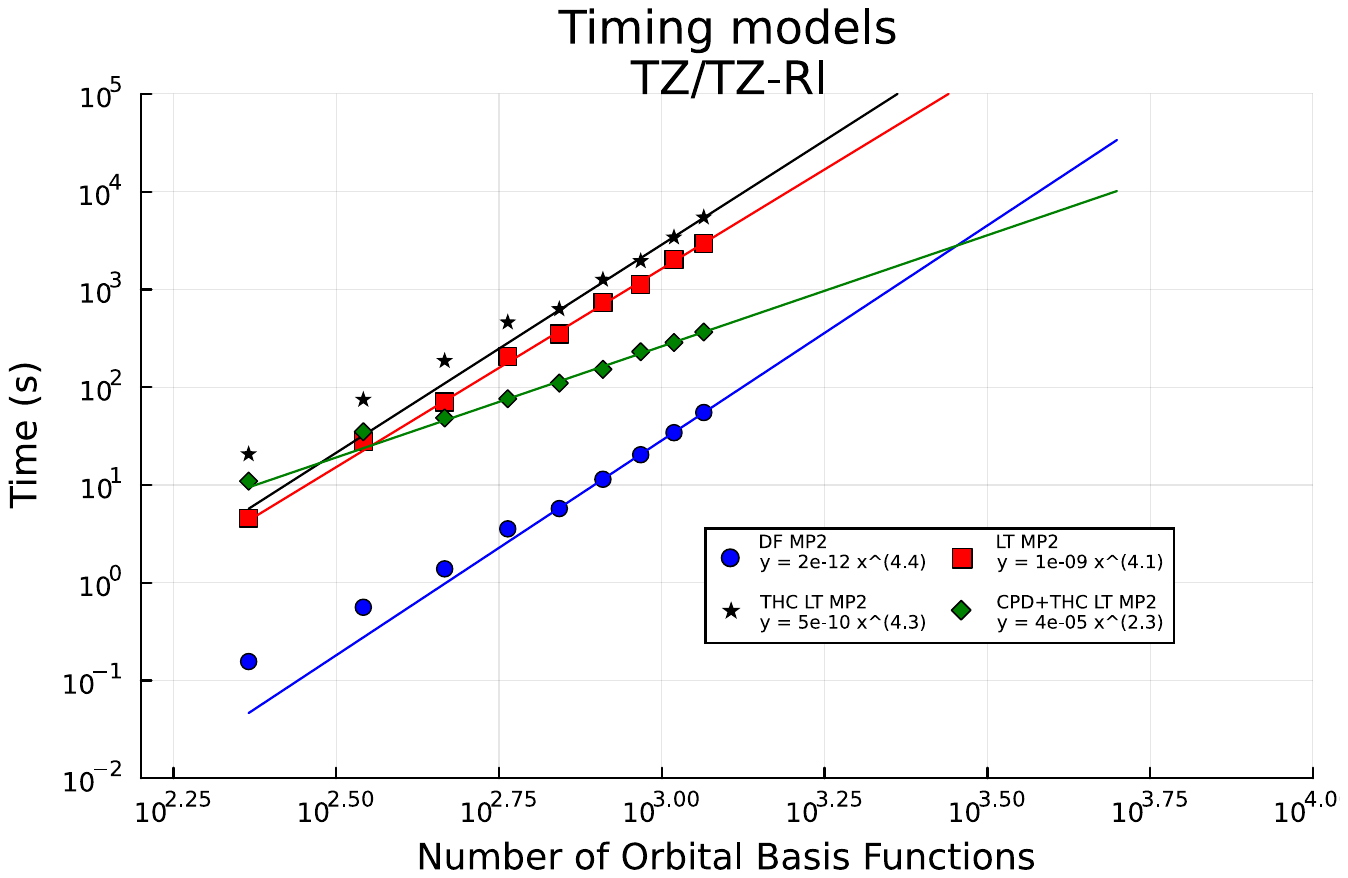}
    \caption{Modeling of the computational effort of different MP2 methods for water clusters with between 4 and 20 molecules in the TZ/TZ-RI basis. The blue circles represent the DF MP2 method, the red squares represent the LT MP2 method, the black stars represent the THC LT MP2 method with $R_\mathrm{THC} = 3X$ and the green diamonds represent the CPD+THC LT MP2 method with $R_\mathrm{CPD}=3X$ and $R_\mathrm{THC}=3X$. All calculations were run using 16 CPU cores}\label{fig:model}
\end{figure}

In \cref{fig:model} we fit the wall-time cost versus orbital basis functions for water clusters with between 4 and 20 molecules in the TZ/TZ-RI basis run on a 16 core partition of the Rome cluster.
In this figure the blue circles represent the canonical DF MP2 method, the red squares represnt the LT MP2 method, the black stars represent the THC approximated LT MP2 with $R_\mathrm{THC}=3X$ and the green diamonds represnt our new CPD+THC LT MP2 method with $R_\mathrm{THC}=3X$ and $R_\mathrm{CPD}=3X$. 
In this figure we represent our CPD+THC LT MP2 method with $t_\mathrm{CPD-ISDF-LT}$ which reflects both the CPD optimization and the MP2 energy computation.
As you can see, our data has a realized scaling reduction over all of the other methods. 
This scaling is smaller than the expected $N^3$ and is most likely a reflection of the small problem sizes.
For example, the largest matrix in this dataset is approximately $0.6$ GB and the largest matrix multiplication requires approximate 0.5 TFLOPs which can be completed in less than 1 seconds on our 16 CPU core partition.
Though it should be noted that this result uses a fixed CP and THC rank however a more thorough study on the scaling of the CP and THC rank with increasing molecular size is necessary to verify the accuracy of these results.
Also in \cref{fig:model}, we extrapolate our model to determine the cross-over between CPD+THC LT MP2 and canonical DF MP2.
The cross-over value is projected to be around 3000 orbital basis functions.
A chemical system with 3,000 orbital functions would require a CPD and THC rank of approximately 20,000 and would therefore generate tensors with storage requirements of about 3GB and smaller. 

\begin{figure}
\begin{subfigure}{0.49\textwidth}
        \includegraphics[width=\linewidth]{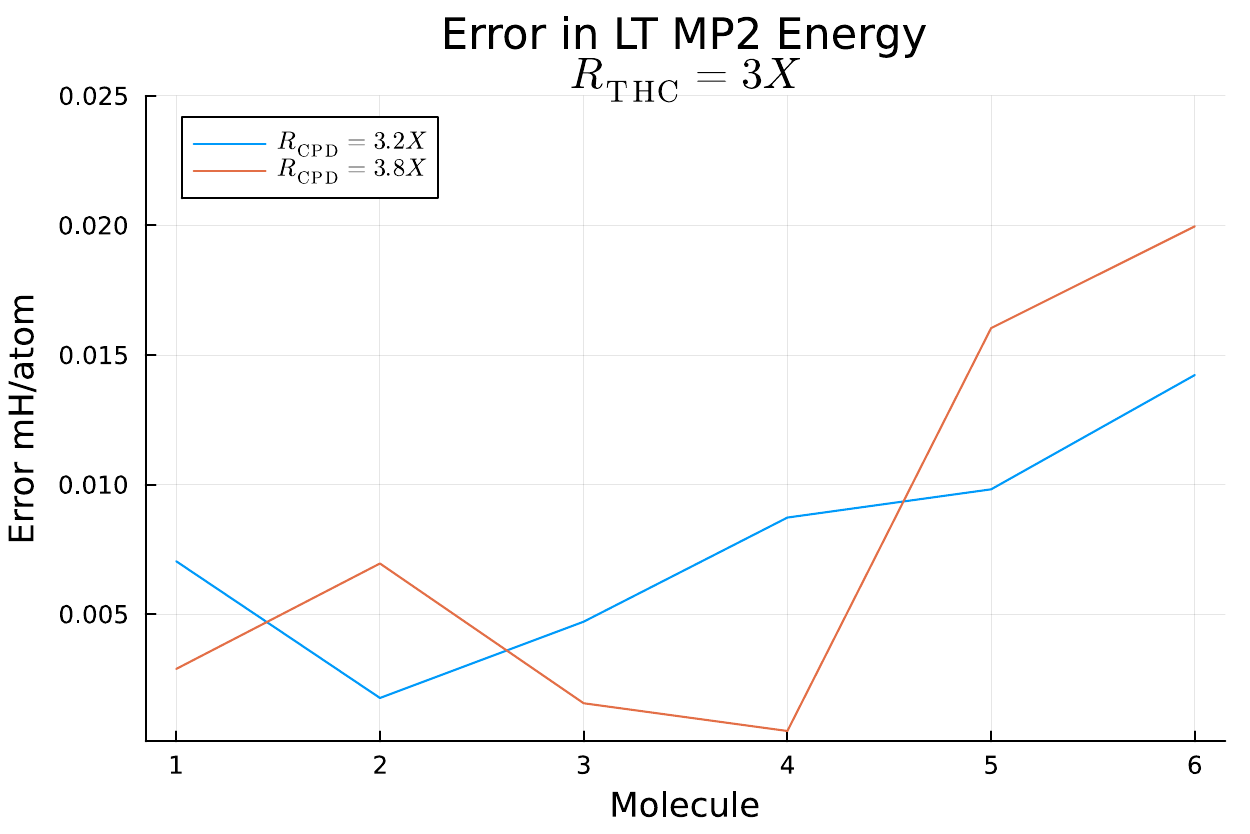}
        \caption{}\label{fig:l7_err}
    \end{subfigure}\hfill
    \begin{subfigure}{0.49\textwidth}
        \includegraphics[width=\linewidth]{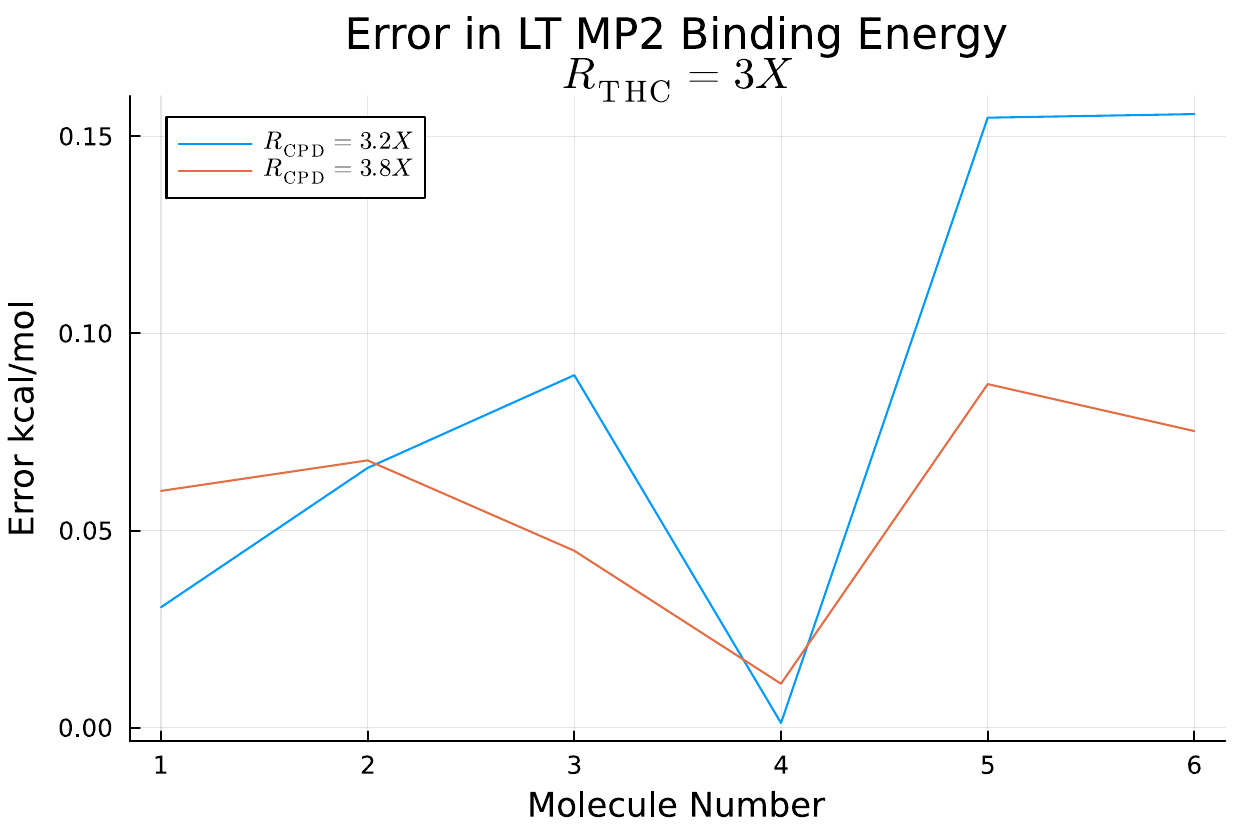}
        \caption{}\label{fig:binding}
    \end{subfigure}
\caption{Error in LT MP2 energy using the CPD+THC LT MP2 method for six of the seven molecules in the L7 dataset with a fixed $R_\mathrm{THC}=3X$. (a) shows the absolute error in energy per non-hydrogen atom and (b) shows the error in the binding energy of the L7 clusters.}\label{fig:L6data}
\end{figure}

Finally in \cref{fig:L6data} we present the error in the CPD+THC LT MP2 method for 6 of the molecules in the L7 set ordered from least to most number of non-hydrogen atoms in the TZ/TZ-RI basis with a fixed THC rank of $3X$ and two different CPD rank $3.2X$ and $3.8X$. 
We omit the largest molecule in this set (the circumcoronene ... GC base pair) because the LT MP2 calculation ran out of memory on our 1 TB node.
Though, the CPD+THC LT MP2 method had no issue with this molecule.
In \cref{fig:l7_err} we show the absolute energy error with respect to the LT MP2 energy.
We see that for both values of CP rank, all calculations introduce an error less than $50$ $\mu H$.
Finally in \cref{fig:binding}, we compute the binding energy error of CPD+THC LT MP2 with respect to the LT MP2 binding energy. 
We see that for the smaller value of $R_\mathrm{CPD}$ the two molecules with the most number of atoms have an error slightly larger than our desired tolerance of $0.1$ kcal/mol. 
However, we are able to correct these values by increasing the CP rank to $3.8X$.
At this larger rank, all studied molecules have a binding energy error of less than $0.1$ kcal/mol.

\section{Conclusions}\label{sec:conclusions}
We introduce a new and efficient method to compute the  CPD of higher-order tensors from their THC approximation.
This method can be combined with efficient THC optimization strategies, such as the the interpolative separable density fitting approximation.
This decomposition strategy can be applied to any THC-like tensor network, to reduce the complexity of computing the order-$d$ CPD from $\mathcal{O}(N^dR_\mathrm{THC})$ to $\mathcal{O}(NR^2_\mathrm{CPD})$ and reduces the storage cost of the CPD optimization from $\mathcal{O}(N^d)$ to $\mathcal{O}(R^2)$, where $R$ is either the THC rank or the CPD rank.
We hope to further accelerate this decomposition strategy by developing GPU implementations and by taking advantage of other mathematical techniques such as randomized linear algebra.

With this reduced-cost CPD optimization algorithm, we study the effectiveness of algorithms which combine the CPD and THC approximation. 
We introduce a combined CPD and THC approximated LT MP2 theory.
In an effort to balance computational efficiency with the accuracy of the THC approach, we only introduce the CPD into the MP2 exchange component.
We find this CPD+THC LT MP2 approach can preserve the accuracy of the LT MP2 method while significantly reducing the wall-time and storage cost of the MP2 approach.
The reduced storage complexity allows for the CPD+THC LT MP2 method to more easily take advantage of GPU accelerators by making it possible to store large problems on the limited GPU memory and therefore minimizing CPU-GPU communications.
In future studies we also hope to integrate this CPD+THC method with embedding and fragmentation approaches to efficiently compute the correlation energy of even larger molecular systems.
Furthermore, we plan to leverage the CPD+THC approximation to efficiently compute MP2 gradients.
The CPD+THC approximation is not limited to the MPn family of methods and we are currently investigating its application to other electronic structure methods for both molecular and periodic systems.

\begin{acknowledgement}
This work was supported by the Flatiron Institute and the Simons foundation. We also acknowledge the Scientific Computing Core (SCC) at Flatiron Institute\\ (https://www.simonsfoundation.org/flatiron/scientific-computing-core/) for providing computational resources and technical support that have contributed to the results reported within this paper. 
\end{acknowledgement}

\bibliography{kmprefs}

\end{document}

%% file: preamble.tex
  \usepackage{geometry} 
  \usepackage{stackengine}
  \usepackage{titlesec}

  \usepackage[T1]{fontenc} 
  \usepackage{amsmath}     
  \usepackage{amssymb}
  \usepackage{bm}
  \usepackage{mathrsfs}
  \usepackage{etoolbox}
  \usepackage{environ}
  \usepackage[version=4]{mhchem}
  \usepackage{booktabs}
  \usepackage{adjustbox}
  \usepackage{hhline}

  \usepackage{setspace}    
  \usepackage{float}       
  \newfloat{chart}{htbp}{loc}
  \newfloat{graph}{htbp}{loh}
  \newfloat{scheme}{htbp}{los}

  \usepackage{graphicx}    

  \usepackage[english]{babel}
  \usepackage{array}

  \usepackage{subcaption}  
  \usepackage{outlines}    
  \usepackage[normalem]{ulem} 

  \usepackage[numbers,sort&compress,super]{natbib}
  \usepackage{natmove}

  \usepackage{cancel}      
  \usepackage{pdfpages}  
  \usepackage[font=scriptsize,labelfont=bf]{caption}

  \usepackage{hyperref} 
  \usepackage{cleveref}	
  	\crefname{figure}{Figure}{Figures}
  	\crefname{table}{Table}{Tables}
  	\crefname{equation}{Eq.}{Eqs.}
  	\crefname{section}{Section}{Sections}
  	\crefname{subsection}{Section}{Sections}
  	\crefname{subsubsection}{Section}{Sections}
  	\crefname{algorithm}{Algorithm}{Algorithms}

  \usepackage{tikz}

    \usepackage[ruled,vlined]{algorithm2e}

  \usepackage{todonotes}

\usepackage{epstopdf}
\AppendGraphicsExtensions{.tiff}